%% file: ms.tex
\documentclass[twocolumn]{aastex7} 

\usepackage{graphicx}
\usepackage{hyperref}
\usepackage{float}
\usepackage{enumitem}
\usepackage{longtable}
\begin{document}

\title{V347 Aurigae: from Peak to Quiescence}

\author[0009-0004-1721-4803]{Diya T. Kumar}
\affiliation{Department of Astronomy, MC 249-17, California Institute of Technology, Pasadena, CA 91125; USA}
\email{diya@caltech.edu}

\author{Lynne A. Hillenbrand}
\affiliation{Department of Astronomy, MC 249-17, California Institute of Technology, Pasadena, CA 91125; USA}
\email{lah@astro.caltech.edu}
\correspondingauthor{Lynne A. Hillenbrand}

\author[0000-0002-0531-1073]{Howard Isaacson}
\affiliation{Astronomy Department, University of California, Berkeley, CA 94720, USA}
\email{hisaacson@berkeley.edu}

 \begin{abstract}
We present an analysis of eight epochs of high-resolution optical spectra of the YSO 
 (young stellar object) V347 Aur.  The source exhibits periodic brightness variations 
 spanning $\sim$2 mag over several months, interpreted as cyclic accretion-burst and quiescent phases, 
 {making V347 Aur a useful probe of the broad range of accretion and outflow
 variability in YSOs.} 
 When faint, V347 Aur shows an M-type photospheric absorption spectrum; 
 when bright, it exhibits a strong emission-line spectrum. 
 Our Keck/HIRES spectroscopy was obtained during the decay of an accretion burst.
 The emission spectrum {includes the strong H I, Ca II, and He I transitions commonly seen in YSOs,
 as well as numerous weaker metal lines, for which we provide a complete inventory.}
 As the star fades, photospheric absorption lines deepen, consistent with decreasing veiling
 from the accretion continuum. H I and He I emission, which trace accretion,
 strengthen relative to the continuum as it fades, {but show the expected decline
 in strength with decreasing accretion after correcting for the strong early veiling.}
 Ca II behaves similarly, with line profiles that alternate between red- and blueshifted dominance,
 which we interpret as the rotation of the magnetospheric accretion flows.
 Narrow, slightly blueshifted absorption in Ca II, Na I, and K I indicates disk winds
 with localized density enhancements. Forbidden [O I] and [S II] emission is blueshifted,
 with broad shoulders and low- and high-velocity narrow peaks,
 consistent with a disk geometry that obscures much of the receding outflow.
 Their behavior as the continuum fades is similar to that of the broad accretion lines.
{The V347 Aur system is a useful proxy both for the ubiquitous short-term accretion variability observed in YSOs, and for their secular evolution on longer time scales.}
 \end{abstract}


\section{Introduction}
 Young stellar objects (YSOs) are early-phase pre-main-sequence stars that often exhibit strong photometric and spectroscopic variability. This variability can arise from multiple physical processes including stellar rotation, surface magnetic activity, and most dramatically, unsteady accretion from a circumstellar disk onto the stellar surface \citep{fischer2023}. 
 The accretion flow is channeled from the inner disk to the star via the magnetosphere \citep{koenigl1991}. 
{Empirically, there are correlated changes in accreting YSOs among photometric brightness,
spectral lines that are diagnostic of disk-to-star accretion, and spectral lines that
measure the strength and kinematics of outflowing winds and jets \citep{edwards1987,fang2023, babnina2026}.}
 In some YSOs, accretion occurs in bursts or episodes, with variable accretion rates that are driven by instabilities in the magnetosphere or the inner disk \citep{romanova2008,dangelo2010}, or further out in the disk due to interactions with orbiting companions \citep{reipurth2000, kuruwita2020}. These episodic accretion events can dramatically affect the observed source brightness, the spectral line profiles, and the continuum veiling.

 V347 Aurigae is a highly variable, pre-main sequence star located in the small, isolated dark cloud L1438, about $200$ pc away. 
 \citet{flores2024} determined stellar parameters of $T_{eff}= 3200$ K, log g = 2.9 dex cm~s$^{-2}$, v~sin$i$=12 km~s$^{-1}$, and $B_{phot}$ = 1.3 kG.
 {\citet{donati2024} derived similar values of $T_{eff}=3300$ K, log g = 3.3 dex cm~s$^{-2}$, v~sin$i$=12 km~s$^{-1}$, and also analyzed the source relative to stellar evolutionary models
 to find $M_{*} = 0.33~M_\odot$ and $R_{*} = 2.2~R_\odot$ with an inferred stellar age $<1$ Myr.
 }
 Long-term archival photometry \citep{dahm2020} revealed periodic brightness variations having large amplitude, about 2 magnitudes in the V band, that occur on a timescale of $\approx 160$ days, 
 and that have persisted for decades.
 The star also illuminates a small reflection nebula that brightens and fades in coordination with the star's photometric variability. 

 Spectroscopic observations \citep{dahm2020} showed that during its photometric bright phases, V347 Aur displays an emission line spectrum that is dominated by strong HI Balmer, CaII, and HeI lines, weaker emission from FeI and FeII lines, and forbidden-lines of [OI] and [SII]. Absorption components are present in the wind-diagnostic lines of KI and NaD. In its fainter phases, the spectrum is dominated by absorption features from the underlying M-type photosphere. The presence of strong veiling during outburst phases, and the large variability in observed line strength and profile morphology, suggest large changes in the accretion rate between the faint and bright states.
 {\citet{giannini2026} recently characterized low-resolution spectra of V347 Aur 
 during both  lower and higher accretion states. 
 Their Figure 2 shows a strong emission-line spectrum in the blue optical 
 even in a relatively quiescent accretion phase having $\dot{M}=10^{-7.71}~M_\odot$/yr, 
 compared to to $10^{-6.80}~M_\odot$/yr during an active outburst. 
 \citet{donati2024} sampled a broader range of accretion in V347 Aur,
 finding rates from $10^{-9.0}$ to $10^{-6.2}$ $~M_\odot$/yr. }
 
 Dahm and Hillenbrand proposed that the recurring outbursts of V347 Aur could be in response to a companion or disk structure that periodically enhances the accretion rate. 
 However, no companion was detected to the limits of the spectra under their study.
 \cite{donati2024} present the results of a 3.5 year monitoring campaign of V347 Aur using the SPIRou spectropolarimeter at CFHT and report that the source is in fact an eccentric binary system,
 refining the period value to $154 \pm 0.7$ days. 
 These authors provide evidence for a brown dwarf companion having a mass of $29 \pm 1.6~M_{\rm Jup}$ and an eccentricity of $0.28 \pm 0.12$, confirming the earlier hypothesis by Dahm and Hillenbrand that a companion may be responsible for triggering the observed periodic accretion bursts. The intensity of emission lines was found to vary significantly, with episodes of strong veiling detected during bright phases. \cite{donati2024} also found that strong to extreme accretion events were occurring near the orbit periastron. 

 In this paper, we present an analysis of high-resolution optical spectra collected with Keck/HIRES for V347 Aur.  The data were obtained over a time span of approximately 3 months relative to the $\sim$5 month full cycle. The spectra capture the system during its decline from a photometric maximum into quiescence. By tracking the time evolution of the spectral lines, we characterize how the veiling, accretion flow, and wind structure evolve throughout the outburst. The changes we see are presumed to measure a decrease in the disk-to-star accretion rate, the accompanying cooling of the accretion shock at the stellar surface, and a decrease in the accretion-driven wind. 

 {Studying V347 Aurigae in this detail provides an opportunity to investigate a system that contains several unusual properties. First, V347 Aurigae is an M-type pre-main star that is completely isolated \citep{dahm2020}, representing a different environment than the clusters in which most stars form. Second, the presence of a close brown dwarf companion represents a relatively uncommon configuration \citep{marcy&butler2000, artigau2021}. Third, its periodic and sustained outbursts make V347 Aurigae a valuable system to study since it can be monitored over complete cycles that sample high to low accretion states. 
 The connection between the photometric evolution of the outbursts and the emission-line spectrum has not yet been explored, however. Our time-series spectroscopy covers  the decline from photometric maximum and thus allow us to track the evolution of individual spectral features while distinguishing changes in veiling from the line-profile behavior. Mapping these spectral changes across the order-of-magnitude changes in the accretion rate can inform our understanding of the accretion-dependent properties of YSOs more generally.  The V347 Aur system is a useful proxy both for the ubiquitous short-term accretion variability seen in YSOs, in which the changes exhibited by different objects range over several orders of magnitude in their amplitudes and in their time scales, and for the general case of accretion and outflow in YSOs, where mass flow rates among large samples span ~1.5 dex at any given stellar mass and age.}

 In Section 2 we present the source light curve and describe our spectroscopic dataset. In Section 3 we present our methods in characterizing the spectral changes over time, followed in Section 4 by the results of our line-by-line analysis. Section 5 provides a discussion of the findings, and our summary appears in Section 6. 

\section{Data}

\subsection{Light Curve}
\begin{figure}
    \centering
    \includegraphics[width=\linewidth]{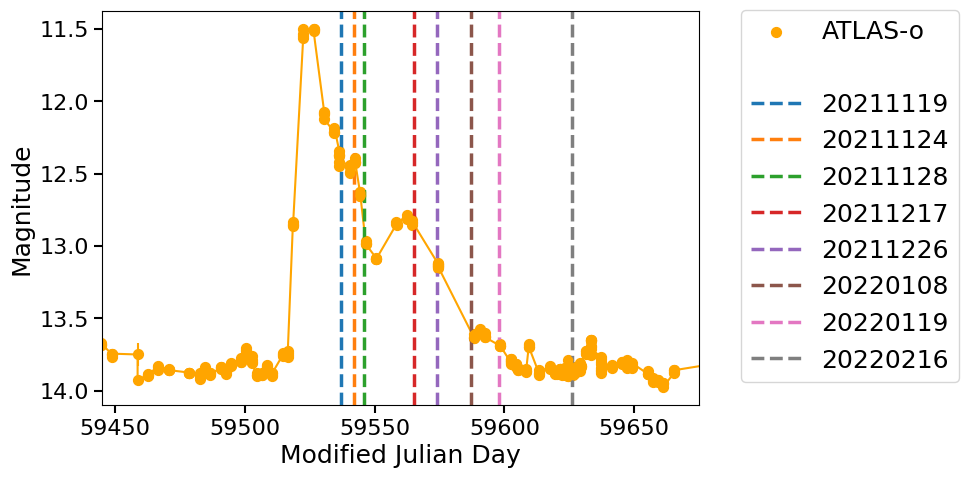}
    \caption{Light curve for V347 Aur obtained from the ATLAS survey  \citep{tonry2018}. Photometry is in the ``orange" band spanning 560-820 nm.
    Dashed lines mark the dates on which the Keck/HIRES spectra were collected.  For ease of comparison, the line colors are consistent with those used in the spectral plots presented below.
    }
    \label{light_curve}
\end{figure}

 An optical light curve of V347 Aur is shown in Figure~\ref{light_curve}. 
 The photometric observations are from the ATLAS survey \citep{tonry2018}, 
 truncated to one full burst cycle of the source {that contains our spectroscopic time series}.
 The dates of our high dispersion spectroscopic observations are marked. 

\begin{table*}[hb]
\centering
\caption{{Observation Dates and Ancillary Information}.}
\resizebox{\textwidth}{!}{
\begin{tabular}{|c|c|c|c|c|c|c|c|}
\hline

\textbf{MJD} & \textbf{Calendar Date} & \textbf {Days Past Peak} & \textbf{Magnitude} & \textbf{S/N } & \textbf{S/N } & \textbf{S/N} & \textbf{Corrections to Heliocentric} \\
\textbf{} & \textbf{} & \textbf{} & \textbf{}  & \textbf{(4605{\AA}-4615{\AA})} & \textbf{(5545{\AA}-5558{\AA})} & \textbf{(7540{\AA}-7550{\AA})} & \textbf{Reference Frame (km/s)} \\
\hline
59537 & November 19, 2021 & 10.6 & 12.423 & 15.3 & 32.5 & 70.6 & ~+9.66 \\
59542 & November 24, 2021 & 15.6 & 12.416 & 15.4 & 28.5 & 70.5 & ~+7.49 \\ 
59546 & November 28, 2021 & 19.6 & 12.893 & 5.1 & 18.8 & 48.1 & ~+6.90 \\
59565 & December 17, 2021 & 38.6 & 12.848 & 8.9 & 23.4 & 61.2 & ~-3.06 \\
59574 & December 26, 2021 & 47.6 & 13.111 & 8 & 19.7 & 51.9 & ~-7.14 \\
59587 & January 8, 2022 & 60.6 & 13.589 & 6.2 & 15.6 & 44.2 & -12.83 \\
59598 & January 19, 2022 & 71.6 & 13.687 & 2.9 & 11.7 & 40.2 & -16.89 \\
59626 & February 16, 2022 & 99.6 &13.876& 3 & 3.8 & 30.7 & -24.58 \\
\hline
\end{tabular}%
}
\label{tab:snr_values}
\end{table*}

\begin{table*}[hb]
\centering
\caption{{Equivalent Widths in \AA\ for Measured Spectral Lines} and Continuum Veiling (r) Values.}
\resizebox{\textwidth}{!}{
\Large
\begin{tabular}{l|rrrrrrrrrrrr}
\hline
\textbf{Calendar Date} & \textbf{\ion{Li}{1}} & \textbf{H$\alpha$} & \textbf{H$\gamma$} & \textbf{H$\delta$} & \textbf{CaII (3968.47} & \textbf{CaII (3933.66)} & \textbf{HeI (5875.65)} & \textbf{HeI (6678.15)} & \textbf{FeI} & \textbf{FeII} &
\textbf{$r_{6707}$} & \textbf{$r_{7100}$} 
\\
\hline
November 19, 2021 & 
$0.006 \pm 0.056$ & $-59.916 \pm 4.593$ & $-11.832 \pm 0.091$ & $-6.954 \pm 0.912$ & $-14.295 \pm 0.956$ & $-16.999 \pm 3.729$ & $ -1.676 \pm 0.168$ & $-1.329 \pm 0.150$ 
& $-1.864 \pm 0.125$ & $-7.587 \pm 0.087$
&13 & 8.8
\\
November 24, 2021 & 
$0.031 \pm 0.078$ & $-60.402 \pm 4.699$ & $-12.325 \pm 0.127$ & $-0.511 \pm 0.545$ & $-16.780 \pm 1.403$ & $ -19.220 \pm 5.248$ & $-2.307 \pm 0.193$ & $-1.468 \pm 0.149$ & $-1.847 \pm 0.157$ & $-8.127 \pm 0.139$
&5.9& 4.95
\\
November 28, 2021 &
$0.055 \pm 0.070$ & $-69.212 \pm 5.698$ &$-21.827 \pm 2.113$ & $-14.536 \pm 1.394$ & $-40.524 \pm 15.020$ & $-104.276 \pm 338.692$ & $-2.895 \pm 0.167$ & $-1.811 \pm 0.127$
& $-2.322 \pm 0.163$  & $-10.258 \pm 0.317$        
&4.8 & 2.6
\\
December 17, 2021 & 
$-0.052 \pm 0.071$ & $-70.221 \pm 6.771$ & $ -19.397 \pm 1.098$ & $-11.338 \pm 0.099$ & $-17.331 \pm 1.997$ & $-21.579 \pm 8.555$ & $-2.299 \pm 0.190$ & $-1.746 \pm 0.146$
& $-2.246 \pm 0.132$ & $-10.165 \pm 0.406$
&$>$15 & 5.0
\\
December 26, 2021 & 
$0.071 \pm 0.093$  & $-78.472 \pm 6.817$ & $-22.649 \pm 1.791$ & $-18.173 \pm 2.180$ & $-38.550 \pm 8.575$ & $-38.973 \pm 22.147$ & $-3.687 \pm 0.165$ & $-1.711 \pm 0.202$
& $-1.673 \pm 0.250$ & $-9.121 \pm 0.234$
&2.6& 2.05
\\
January 8, 2022 &
$0.197 \pm 0.138$ & $-75.837 \pm 5.656$ & $ -24.418 \pm 3.017$ & $-17.502 \pm 1.839$ & $-46.186 \pm 13.193$ & $-40.003 \pm 22.484$ & $-3.057 \pm 0.242$ & $-1.329 \pm 0.316$ 
& $-1.448 \pm 0.364$ & $-7.505 \pm 0.084$
&0.8& 0.85
\\
January 19, 2022 & 
$0.255 \pm 0.138$ & $-80.428 \pm 5.579$ & $-54.632 \pm 16.015$ & $-27.818 \pm 7.160$ & $-59.309 \pm 18.396$ & $-65.162 \pm 49.029$ & $-4.034 \pm 0.178$ & $-1.225 \pm 0.346$ 
& $ -0.649 \pm 0.485$ & $-4.726 \pm 0.135$
&0.4& 0.30
\\
February 16, 2022 &
$0.383 \pm 0.170$ & $-64.451 \pm 3.982$ & N/A & N/A & $-15.171 \pm 3.804$ & $-18.401 \pm 9.938$ & $-2.977 \pm 0.444$ & $ -0.833 \pm 0.405$ 
& $-0.824 \pm 1.745$ &$-3.958 \pm 0.742$
&0.0& 0.0
\\
\hline
\label{equivalent widths}
\end{tabular}
}

\label{tab:veiling_param}
\end{table*}

\subsection{High Dispersion Spectroscopy}
The Keck/HIRES \citep{vogt1994} spectroscopic data set 
{is available through the Keck Observatory Archive} 
and was obtained {using the infrastructure of the California Planet Search}\footnote{https://california-planet-search.github.io/}. This includes observations, image processing, spectral extraction to one-dimensional format, and wavelength calibration,
all as described in \cite{howard2010}.
HIRES provides broad optical coverage across three detectors. In the utilized setup, the blue $b$ chip spans $\sim$3600-4800\AA, the red $r$ chip $\sim$5000-6430\AA, and the ``infrared" $i$ chip $\sim$6500-7990\AA.
The achieved spectral resolution R$\approx$60,000 ($\Delta v \approx 5$ km/s). 

The spectra were obtained on eight separate nights between November 19, 2021, and February 16, 2022, capturing the star at various stages during its decline from outburst (see Figure~\ref{light_curve}). The collection times span the post-outburst fading, including a brief re-brightening between the first and second epochs, a subsequent local maximum during the decline, the continued decline into quiescence, and finally in the quiescent state itself. This temporal coverage enables a detailed view of the evolving spectral line behavior across a factor of nearly 4 in brightness (from 12.42 mag to 13.88 mag).

Typical integration times were 300 seconds and resulted in
signal-to-noise ratios ranging from $\sim$3-15 in $b$, $\sim$4-33 in $r$, and $\sim$31-71 in CCD $i$.  
Table \ref{tab:snr_values} summarizes the observational details, including the corresponding source magnitudes, the realized signal-to-noise ratios across the three CCD bands, and the barycentric velocity corrections for each epoch.

\section{Methods}

\subsection{Spectral Pre-Processing}
 The extracted spectra are stored in a FITS format, with wavelength and flux values read using the \textit{Astropy} package. 
Each of our eight spectra have their own wavelength calibration, thus for a fixed order, individual spectra can exhibit small differences in wavelength coverage and sampling. We therefore interpolated each spectrum onto a common wavelength grid that covers the overlapping wavelength range, resulting in a uniform wavelength spacing (\AA\ per pixel) across the order.

After interpolation to the common wavelength grid, the following set of processing steps 
were applied to clean and standardize the data: outlier removal, smoothing, continuum fitting, and velocity corrections. 

To remove outliers, flux values deviating beyond ten standard deviations from the median {in each order} were masked. 
For visual clarity in plotting, a gaussian filter was applied to some of the lower signal-to-noise spectra at later epochs (see Table \ref{tab:snr_values}) using the \textit{scipy.ndimage.gaussian\_filter1d} routine. 

To normalize the spectra, individual spectra were divided by the observed flux in a fitted continuum around spectral lines of interest. For spectral regions where the continuum showed a noticeable slope, a first order polynomial was used. Orders with a relatively flat continuum were fit with a zeroth order polynomial. The choice of polynomial order was based on the visual inspection of the shape of the baseline.   

To center spectral lines in the stellocentric reference frame, we applied velocity corrections to the individual spectra that take them from the rest frame of the telescope to the rest frame of the star. We have accounted for both the barycentric corrections tabulated in Table \ref{tab:snr_values}, and the radial velocity of the source relative to the Sun. For latter, we have applied a correction of -8.9 km/s, adopting the +8.9 km/s radial velocity found from cross correlation by \cite{dahm2020}, which is consistent with the +8.6 km/s median radial velocity reported by \cite{donati2024} based on LSD analysis. 

We validated the corrections by examining isolated atomic absorption lines in our spectra. 
These are sparse in number, given the heavy veiling for the earlier epochs 
and the cool photosphere dominated by TiO that is seen in the last few epochs.  
In the reddest wavelengths of our spectra, we could identify several plausible lines 
(\ion{Ca}{1} 7326.15 \AA, \ion{Fe}{1} 7401.69 \AA, \ion{Cr}{1} 7400.18 \AA, and \ion{Rb}{1} 7800.27 \AA)
for this purpose, detected only in the later epochs.  The most useful for our purpose is 
the \ion{Rb}{1} line, which is a member of a resonance doublet
(the other line at 7947.60 \AA\ is contaminated by TiO) 
and easily ionized, so {it is} typically seen only in very late M and L type dwarfs.
\ion{Rb}{1} is also gravity-dependent \citep{mcgovern2004}, perhaps explaining its visibility in 
V347 Aur at an earlier spectral type (M4) and young age.
Although weak, with equivalent width only $\sim$0.1 A, \ion{Rb}{1} 7800.27 reaches a depth 
of 25\% below the continuum, and the line is perfectly centered 
in velocity space -- confirming our {heliocentric and radial velocity} corrections.

\subsection{Spectral Analysis}

To identify spectral lines, we use the \textit{find\_peaks} function from the \textit{scipy.signal} module, which detects local maxima in the flux values. The peaks were identified using thresholds on prominence and minimum distance, to ensure that only significant spectral features that were well separated from neighboring maxima were classified as true emission lines. For this analysis, we use the spectrum collected on 2021-12-17. This particular spectrum does not have the highest signal-to-noise ratio, but it provides a particularly favorable view of the emission line inventory, as it corresponds to a local maximum in the light curve. 

A tabulation of detected emission lines appears in Appendix~\ref{sec:lines}.
In the analysis below we focus on a few of the most prominent emission lines. 
However, in its outburst phases, V347 Aur exhibits a rich spectrum of emission lines
consisting mainly of weak \ion{Fe}{1} and \ion{Fe}{2} lines,
but also \ion{Ti}{1} and \ion{Ti}{2},  \ion{Cr}{1}, and \ion{Cr}{2},
\ion{Ca}{1}, and \ion{Mg}{1} features.

Equivalent widths were computed using the \textit{specutils.equivalent\_width} function applied to continuum-normalized spectra. We performed a direct numerical integration of the line profile over defined wavelength regions.  

To quantify the errors on the equivalent width measurements, we adapt a formula
from \citep{vollmann2006}.  However, we modify their formula to account for the fact that 
our lines are in emission rather than absorption,
and also are generally much stronger in equivalent width than their typical wavelength interval $\Delta\lambda$.

We also substituted half of the line peak flux for the average spectral line flux. The lines of interest in this work are typically strong, broad and non-gaussian in shape, with average flux that is thus
significantly decreased relative to a gaussian line.

Our modified equation for the equivalent width error is: 
\begin{equation}
\sigma(W_\lambda) = \left| \sqrt{1 + \frac{\overline{F_c}}{\frac{1}{2} \times F_{max}}} \cdot \frac{\Delta\lambda - |W_\lambda|}{S/N} \right|.
\label{eqw_errors}
\end{equation}

{In the interpretation of the declining light curve from outburst peak
(Figure \ref{light_curve})
as a decrease in accretion rate, we can expect that observed equivalent widths 
are affected by changes to the continuum veiling.
Continuum veiling is denoted by the veiling parameter $r$, defined as the ratio of the excess continuum flux to the intrinsic photospheric continuum flux \citep{hartigan1989}. The presence of excess continuum reduces the apparent equivalent widths of photospheric absorption lines such that
$EW_{true} = EW_{observed} \times (1+r)$.
The veiling factor for each epoch was measured using the least veiled epoch as the reference photospheric spectrum. 
We account for changes in continuum veiling by applying this multiplicative correction to the measured equivalent widths.
{We also propagate the $1+r$ factor in the errors.  At early epochs, the total error is dominated
by the large veiling values, while at later epochs the lower signal-to-noise contributes most of the total error.}
}

{As mentioned above, photospheric absorption is prominent in only the last few epochs
of our data set. The high level of veiling during the early epochs 
means that there are few regions of spectrum that are viable for the measurements.  
We fit two regions: the continuum-normalized \ion{Li}{1} 6707 \AA\ line profile,
and the entire order containing the broad TiO 7100 and 7140 \AA\ bands.
The best fitting value of $r$ was chosen for each epoch by minimizing the $\chi^2$ difference 
between a model  consisting of a veiled version of the final epoch, and the observed spectrum. 
The veiling values from the two regions are broadly similar, except at the earliest epochs
when the veiling values are largest, with the \ion{Li}{1} measurements about 50\% higher. We adopted the TiO based values, which we consider more reliable given that they were derived over a wider range of the spectrum. Both sets of measurements are presented in Table \ref{equivalent widths}. 
}

{In applying a veiling correction, we adopt the approximation that veiling is independent of wavelength across the spectral range of our dataset. 
This has been demonstrated for wavelengths $>5500$ \AA\ observationally \citep{basri1990}
and is true theoretically for veiling values up to around 5 \citep{nelissen2023}. 
While veiling may increase towards shorter wavelengths, the limited 
number of absorption lines exhibited by V347 Aur at the earlier epochs
do not allow us to measure a wavelength-dependent veiling correction. 
}

Finally, to analyze the spectral lines in velocity space, we apply the doppler formula. This transformation allows us to use the morphology of spectral lines to track the motions of gas in the system over time.

\subsection{Quantifying the Time Evolution of Spectral Lines}

For each identified spectral line, we analyze its temporal evolution across epochs by measuring several key properties, including line center shifts, line broadening, observed line strength, and line structure. 

Line center shifts are quantified by tracking changes over time in the wavelength of peak flux, which indicate gas motion. Line broadening is characterized using the full width at half maximum (FWHM), which can indicate the thermal motion of the gas and the rotation of the star. Line strength is measured through equivalent width calculations, {corrected for veiling,} that quantify variations in the density of absorbing or emitting atoms. Finally, changes in line structure are visually examined in the wavelength and velocity plots.

\section{Line Analysis Results}

Our analysis of spectral evolution in V347 Aur is separated into {three} categories. 
We first study absorption features, focusing on TiO and \ion{Li}{1} 6707{\AA} which are the most prominent. 
We then study the strong emission lines, including lines of hydrogen, helium, calcium, and iron. {Finally, we study several outflow diagnostics of slow winds and fast jets.}

The specific emission lines for which
we present line profile time series are: 
H$\alpha$, H$\delta$, H$\gamma$, 
HeI 5875.62 {\AA}, HeI 6678.15 {\AA}, 
CaII 3968.47 {\AA}, CaII 3933.66 {\AA}, 
FeI 5455.61 {\AA}, and FeII 5316.62 {\AA} {for the strong (predominantly accretion) lines}; 
and KI 7698.96 {\AA}, 
NaD 5895.94 {\AA}, 
[OI] 6300.30 {\AA}, and [SII] 6730.82 {\AA} {for the wind/jet diagnostics}. 

 \begin{figure}
     \centering
     \includegraphics[width=\linewidth]{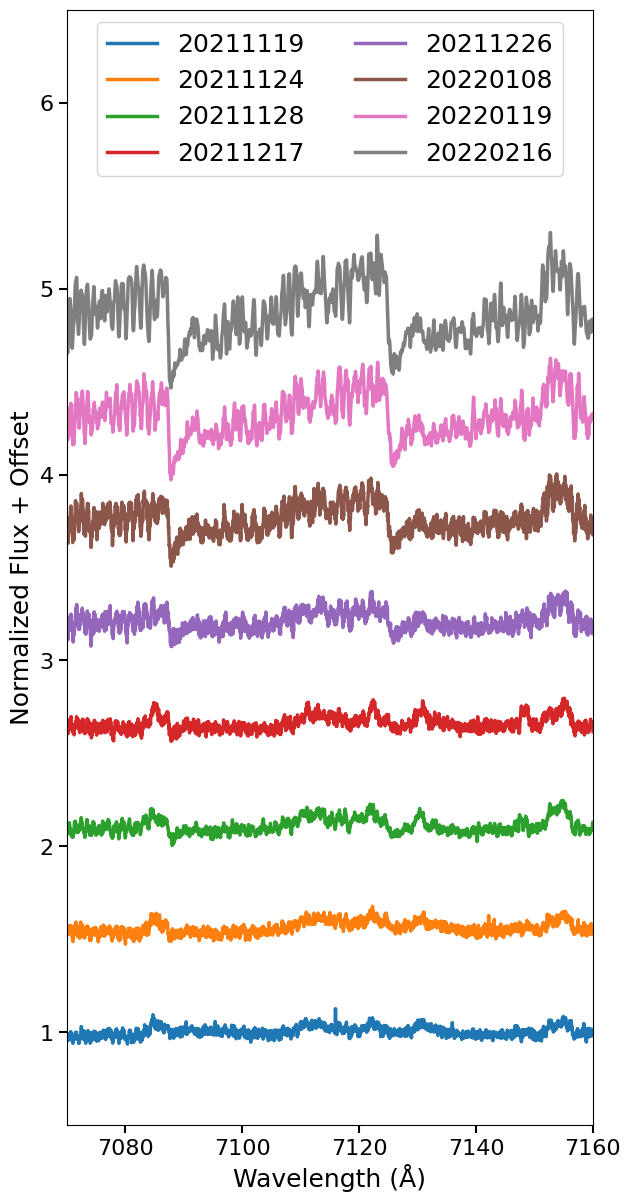}
     \caption{{Spectra normalized to the continuum} in the 7070-7160 {\AA} region, showing the emergence of broad absorption features from TiO as the star V347 Aur fades from an accretion burst. The earlier spectra {towards the bottom of the figure} (e.g., 2021-11-19, 2021-11-24) show a veiled continuum with minimal to no absorption and weak emission lines, consistent with strong accretion activity. In contrast, later spectra {towards the top of the figure} (e.g., 2022-01-08, 2022-02-16) exhibit deep photospheric absorption features, reflecting the decreasing veiling as accretion subsides.
     }
     \label{tioband}
 \end{figure}

\begin{figure*}[ht]
\centering
\includegraphics[width=0.48\linewidth]{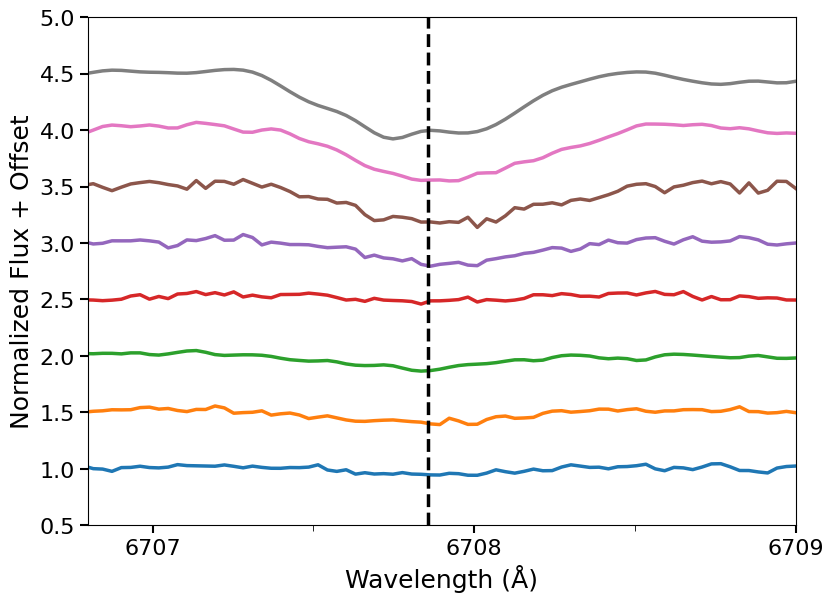}
\includegraphics[width=0.48\linewidth]{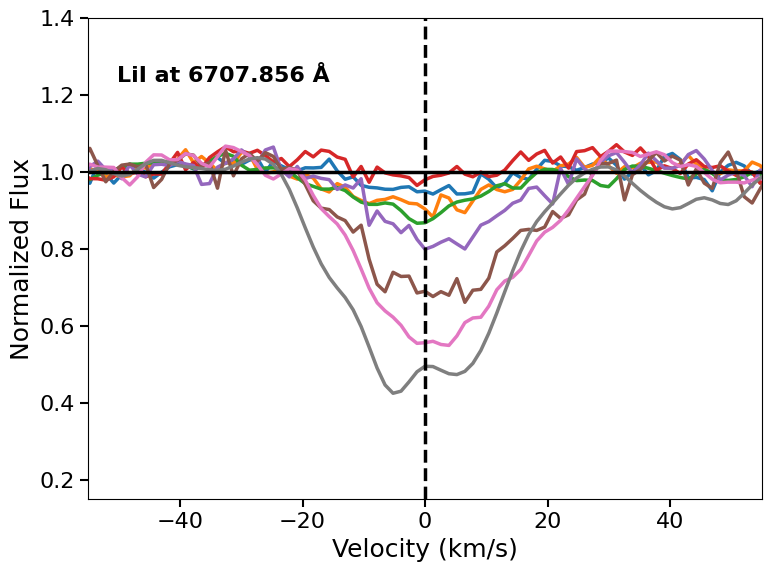}
\\
\includegraphics[width=0.48\linewidth]{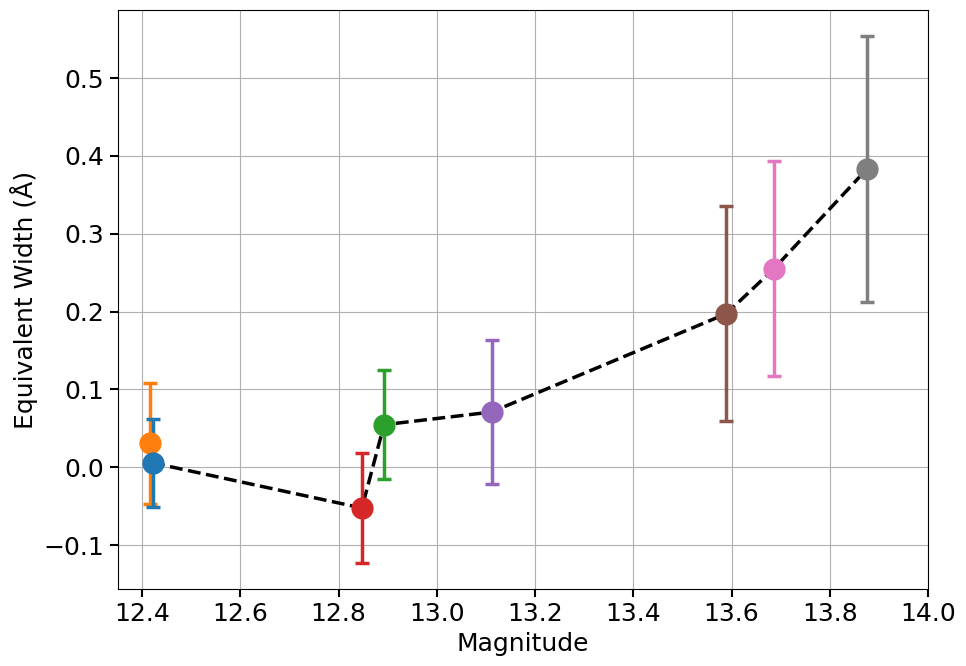}
\includegraphics[width=0.48\linewidth]{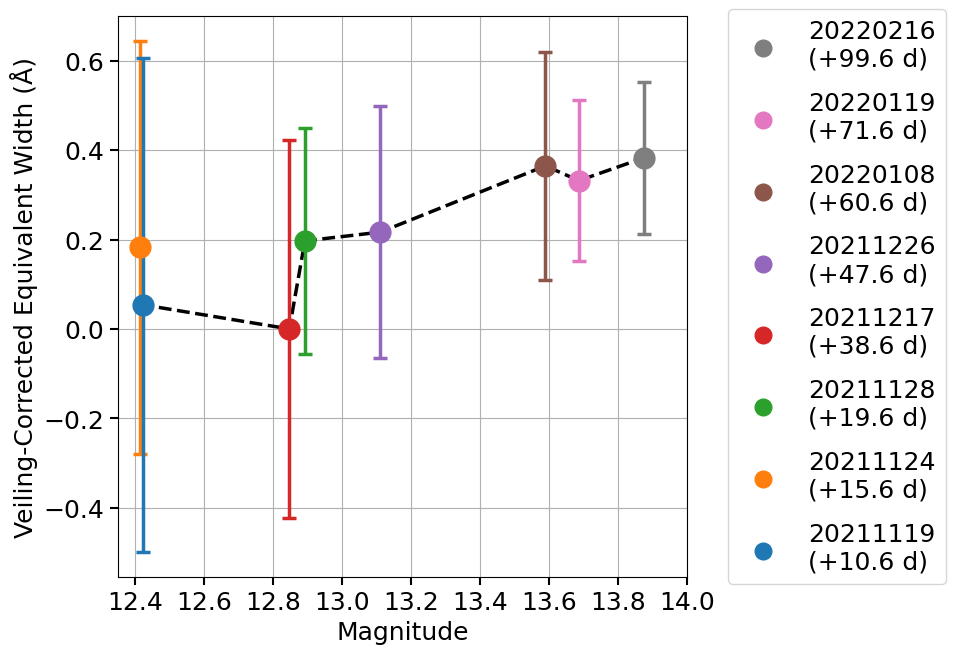}

\caption{\ion{Li}{1} 6707.856 \AA\ line {normalized to continuum (top row), presented} as observed (left) and on a velocity scale (right). Equivalent widths {(bottom row) are presented as measured (left) and veiling corrected (right) {with error propagation}.
The corrected equivalent width at 2021-12-17 is set to zero, since the measured value was 
close to zero but formally negative during this epoch of photometric re-brightening.
The legend includes the number of days since peak brightness.}
As the source fades, the veiling decreases, and the line absorption {deepens}.
}
\label{liI_spectra}
\end{figure*}

\subsection{{Photospheric} Absorption Features}

As V347 Aur fades into quiescence (Figure \ref{light_curve}), absorption features in the spectrum become more pronounced.
{Figure \ref{tioband} illustrates the TiO band region around 7100 \AA}. 
The later-in-time and fainter epochs show spectra with deeper absorption features. These may be compared to the earlier-in-time and brighter epochs when the spectra are more heavily veiled, and there is even weak emission in some atomic lines.

This observation signals a decline in the continuum veiling of spectral absorption lines that is caused by accretion activity. 
During periods of high accretion, excess continuum emission {veils} the absorption lines, making them {appear} shallower. As the accretion rate drops and veiling decreases, the photospheric absorption becomes more apparent.
{As noted above, atomic absorption lines are difficult to see in V347 Aur.
Our analysis below includes only \ion{Li}{1}, which is the most prominent absorption line
aside from the broad features due to TiO.}

\ion{Li}{1} centered at 6707.856 \AA\ is usually a prominent absorption line in YSO spectra. 
This line is expected from theory to be preserved in young stars, for at least some portion of their pre-main sequence lifetimes. 
In these fully convective stars, photospheric lithium is efficiently transported throughout the stellar interior, where it is destroyed. As a result, the strength of lithium serves as an indicator of stellar youth {\citep{JacksonJeffries2014,Soderblom2014,Weymann1963}}.

In V347 Aur, the \ion{Li}{1} feature (Figure \ref{liI_spectra}) apparently strengthens as the star fades. 
The exception to the observed trend is the 2021-12-17 epoch, in which the line is the weakest. This particular spectrum coincides with a small photometric re-brightening (Figure~\ref{light_curve}) and probably reflects a short-lived episode of heightened accretion activity. 
{The \ion{Li}{1} 6707.856 \AA\ line is a doublet,
which even becomes visibly resolved in V347 Aur in the final epoch (gray) as 
the veiling decreases and the profile sharpens.  We note that
in stars with large v~sin\textit{i}, the nearby \ion{Fe}{1} lines at 6704.48 and 6705.11 \AA\ on the blue side and the weaker 6711.30 \AA\ on the red side can blend in and contaminate the observed \ion{Li}{1} profile. However, given V347 Aur's v~sin\textit{i} of $12.0 \pm 0.5$ km s$^{-1}$ \citep{donati2024}, these \ion {Fe}{1} lines are not a significant contaminant here.}
The velocity profile shows that \ion{Li}{1} is consistently centered on the stellar rest velocity across all epochs.  

{To illustrate the effect of the veiling correction, Figure \ref{liI_spectra} displays both the measured and the veiling-corrected equivalent widths. The apparent increase in observed equivalent width towards later epochs is reduced or removed after applying the correction, yielding line strengths that are approximately constant within the uncertainties. This shows that the adopted veiling-correction procedure based on the TiO region recovers the expected instrinsic photospheric line strength, within the errors.}
{We note that for the 2021-12-17 spectrum, the measured equivalent width of the \ion{Li}{1} line is positive, despite the line core being in absorption; this arises because of the broad equivalent width integration range. As emission is unphysical for this line, and inconsistent with visual examination, we have set the corrected equivalent width at this epoch to zero.}

\subsection{{Strong Emission Lines}}
\subsubsection{{Hydrogen Balmer Lines}}

\begin{figure*}
\centering
    \includegraphics[width=0.32\linewidth]{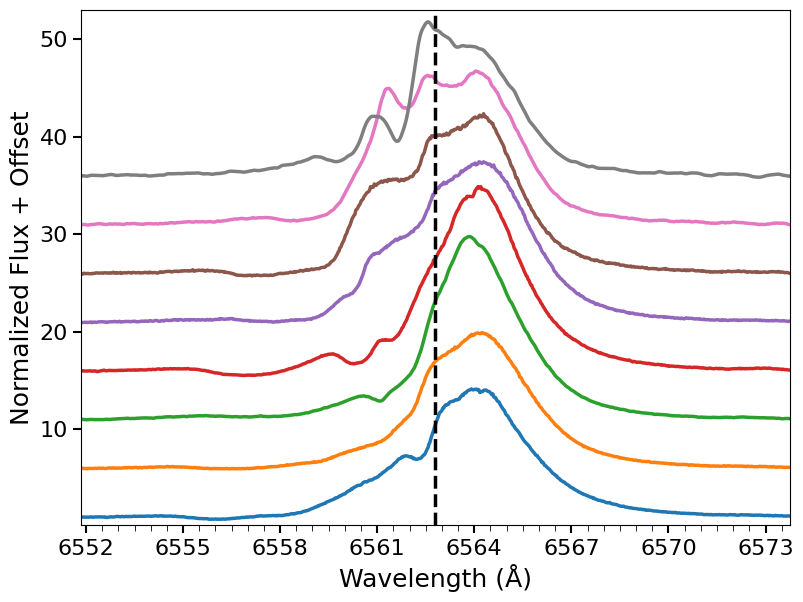}
     \includegraphics[width=0.32\linewidth]{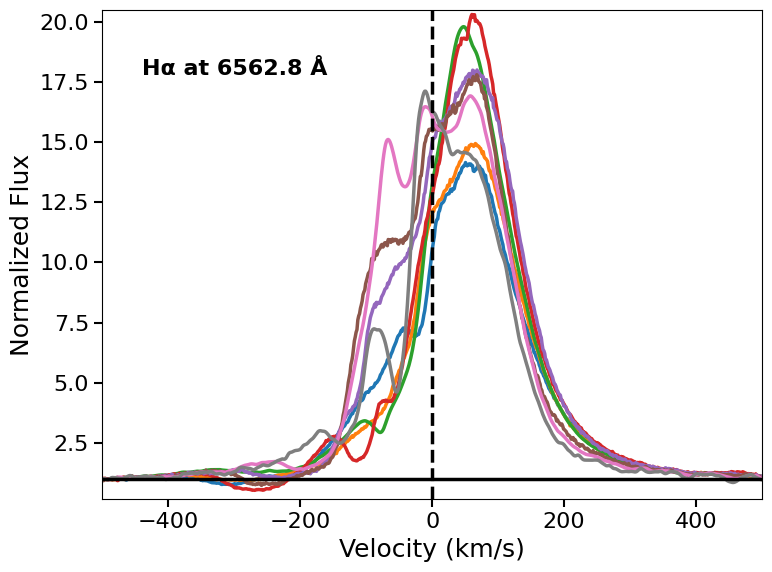}
    \includegraphics[width=0.32\linewidth]{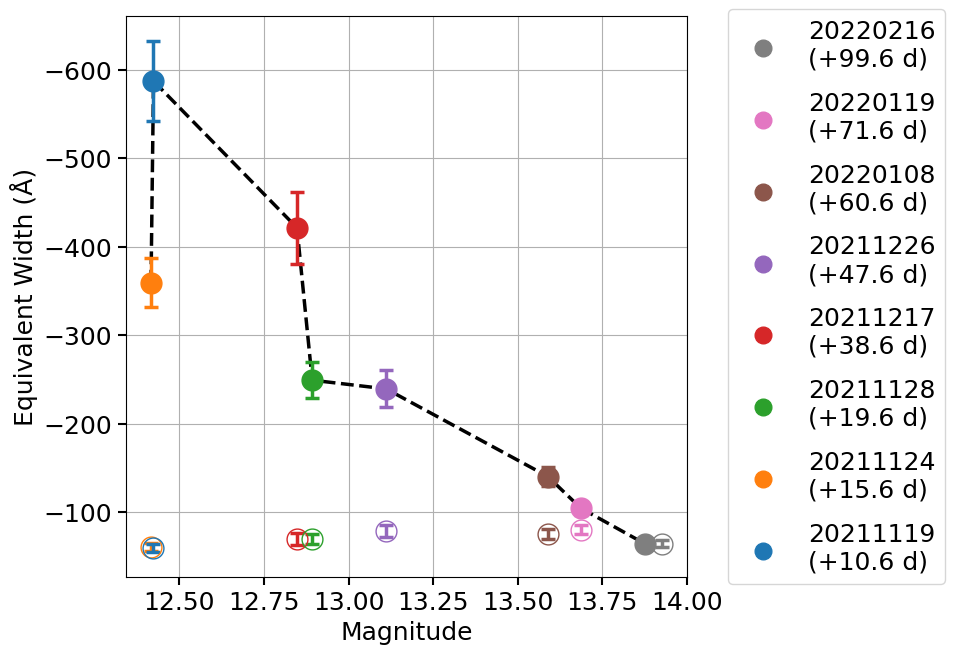} \\

    \includegraphics[width=0.32\linewidth]{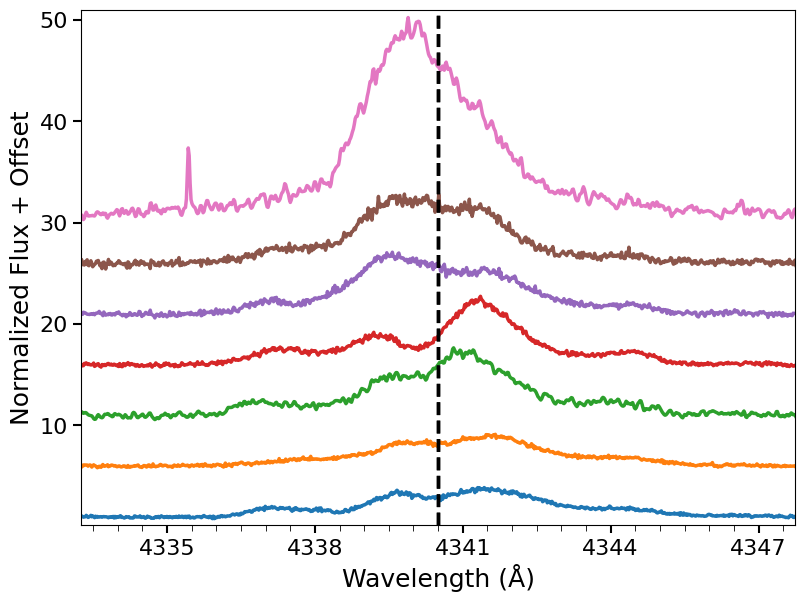}
    \includegraphics[width=0.32\linewidth]{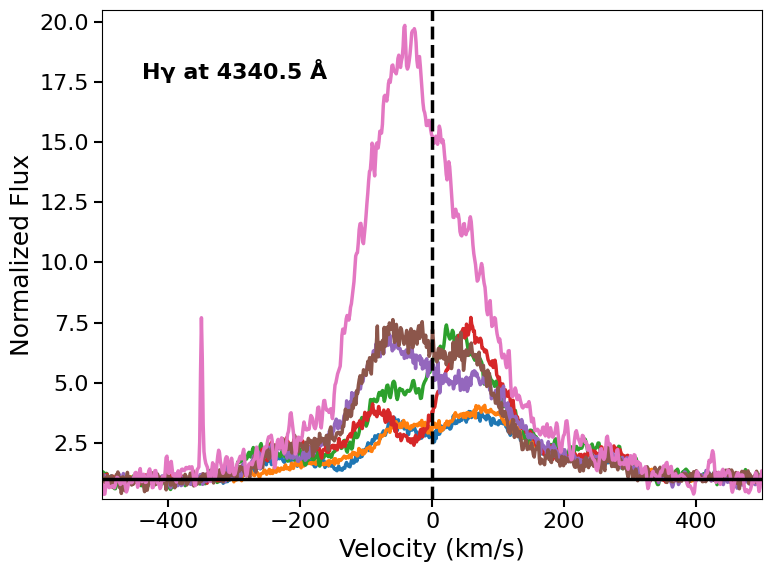} \includegraphics[width=0.32\linewidth]{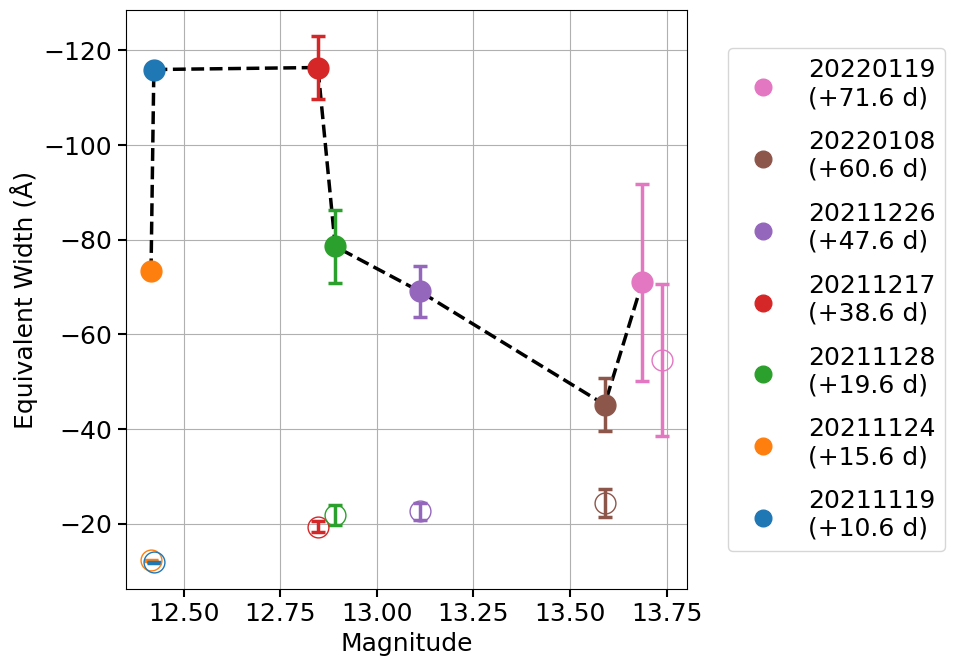} \\  
    \includegraphics[width=0.32\linewidth]{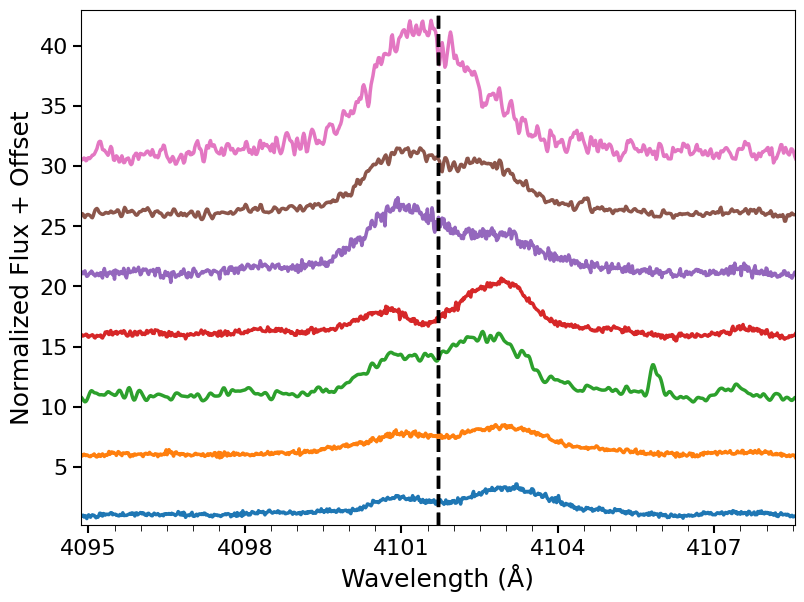}
    \includegraphics[width=0.32\linewidth]{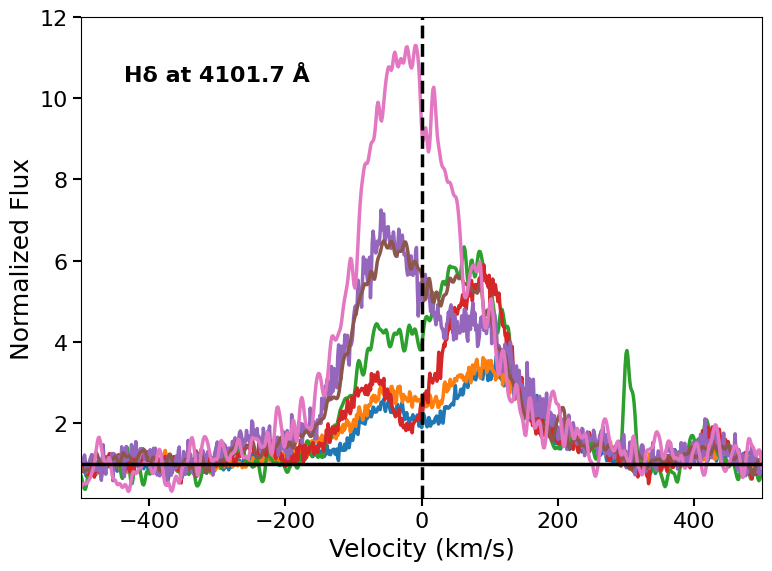}
    \includegraphics[width=0.32\linewidth]{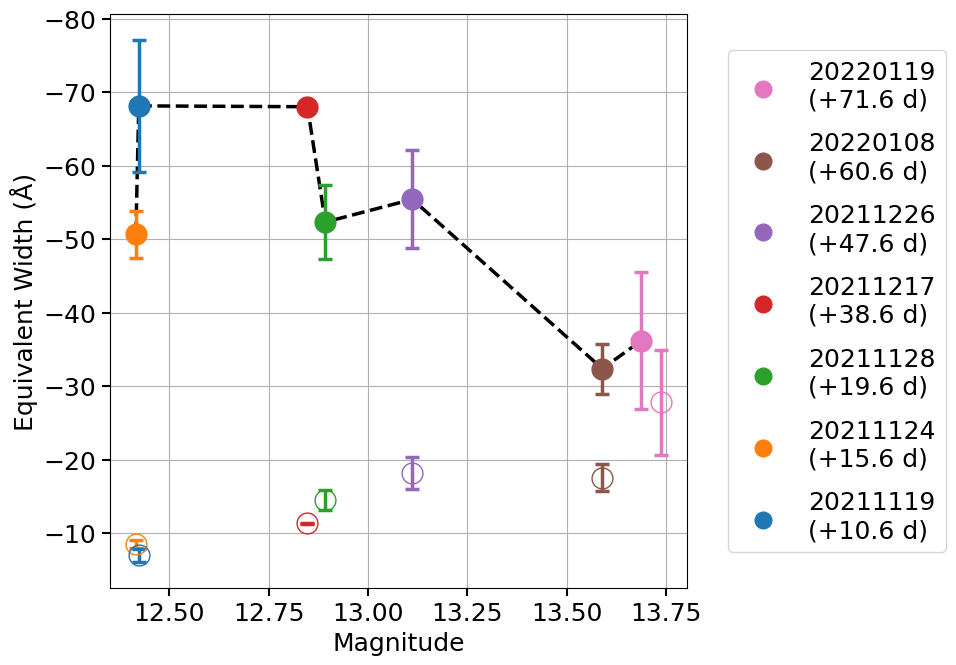} 
    \caption{Spectra around the H-alpha line at 6562.80 \AA\ (top), H-gamma at 4340.50 \AA\ (middle), and H-delta at 4101.70 \AA\ (bottom). Left panels show observed, {continuum-}normalized spectra while middle panels are velocity plots where
    we omit the last spectrum in the H-gamma and H-delta plots due to noise in the continuum. {The right panels show equivalent widths; open circles denote the measured values, while filled circles (connected by a dashed line) denote the veiling-corrected values.} 
    {For visual clarity, the measured (open) equivalent width measurements at later epochs have been offset by 0.05 magnitudes to avoid overlapping error bars with their veiling-corrected (filled) counterparts.}
    As the star fades over time, the spectral lines become broader with enhanced blue-side emission. In addition, some peaks change their morphology into multiple peaks, especially in H-alpha, or are seen to shift in the {location}  of their peaks, for H-gamma and H-delta. The observed line profiles relative to the continuum level generally increase as the star fades, {but the veiling-corrected equivalent widths illustrate that the line strengths are actually decreasing. 
    }
}
    \label{hydrogen_spectra}
\end{figure*}

The Hydrogen Balmer lines are usually strong emission features in spectra of YSOs.  Our V347 Aur data set includes the H$\alpha$, H$\gamma$, and H$\delta$ members of the Balmer series; unfortunately, H$\beta$ falls outside the spectral coverage, between orders on the detector. The Balmer line profiles show signatures of both accreting and outflowing material, with their morphology evolving significantly as V347 Aurigae fades.

As shown in Figure~\ref{hydrogen_spectra}, the Balmer lines broaden and shift to shorter wavelengths as the star fades. 
The evolution of the line profile shape seems to differ between lines.
As seen most prominently 
in H$\alpha$, a distinct asymmetry towards the red in the emission line profiles is present in the earlier observations, while towards the later observations, more blue-side emission becomes apparent. 
The 2021-11-28 spectrum, taken near a photometric local minimum shows mild blueshifting, and {the final four epochs}
show more evidence of it. The last epoch (2022-02-16) is omitted for H$\gamma$ and H$\delta$ due to low signal-to-noise. {As the star fades}, these lines show growth in their blue-side emission relative to the red-side (Figure \ref{hydrogen_spectra}).
The increasing blue-side emission is consistent with the increased visibility of outflowing material. As the accretion decreases, the emission from wind-related components moving towards the observer becomes more apparent. 

The H$\alpha$ profile also has a complex, multi-component structure,  
with various narrow blueshifted absorption that are transiently seen against the broad emission profile.
By the 2021-12-26 spectrum, the blueshifted side of the profile starts to broaden. In the final three epochs, the line exhibits multiple blueshifted peaks, growing to two separate peaks. In contrast, H$\delta$ and H$\gamma$ initially appear as double-peaked profiles with their blueshifted components at -50 km/s and their redshifted components at 100 km/s and 70 km/s respectively (measured from the first collected spectra). Additionally, for H$\gamma$, the redshifted peak is about 1.3 times stronger than the blueshifted one at onset, but the blueshifted peak grows taller than the redshifted one, becoming about 1.5 times stronger. For H$\delta$, the redshifted peak is initially taller than the blueshifted one, by a factor of about 1.16, and the blueshifted peak then grows to about 1.28 times stronger than the redshifted peak. In both of these lines, in the 2022-01-19 spectrum, the double peak disappears and only a single, stronger {blueshifted} peak is present. As noted above, the final epoch for H$\gamma$ and H$\delta$ (2022-02-16) is omitted in the plots due to low signal-to-noise.

 {These variations are further quantified through equivalent width measurements, displayed in Figure \ref{hydrogen_spectra}. In every line, the observed line-to-continuum strengths generally increase as the star gets fainter, but the veiling-corrected equivalent widths generally decrease, as expected. The absolute scale also drops sharply down the Balmer sequence, as expected. We note that rather than a smooth monotonic decline, all three lines show a partial rebound at the 2021-12-17 epoch, likely linked to the brief photometric re-brightening at that time.}

\subsubsection{{Calcium H\&K Lines}}

\begin{figure*} 
\centering
    \includegraphics[width=0.32\linewidth]{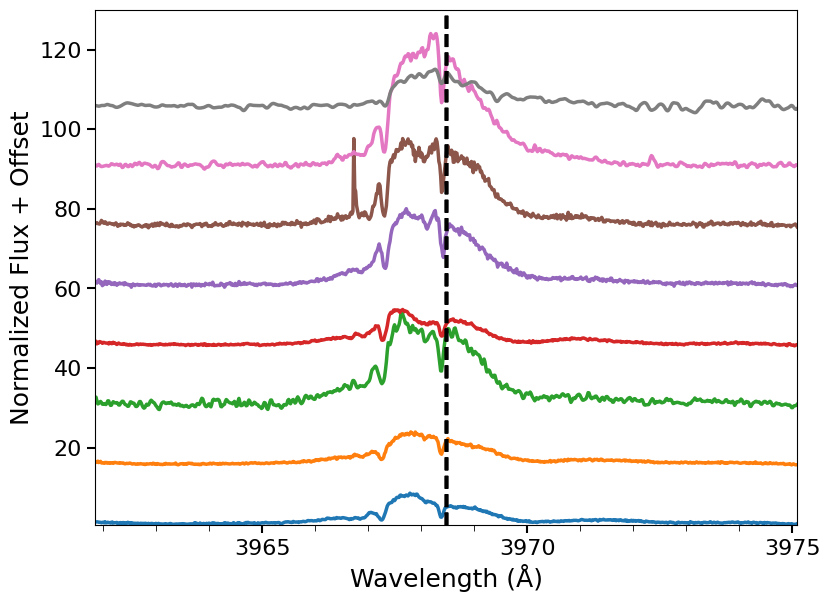}
    \includegraphics[width=0.32\linewidth]{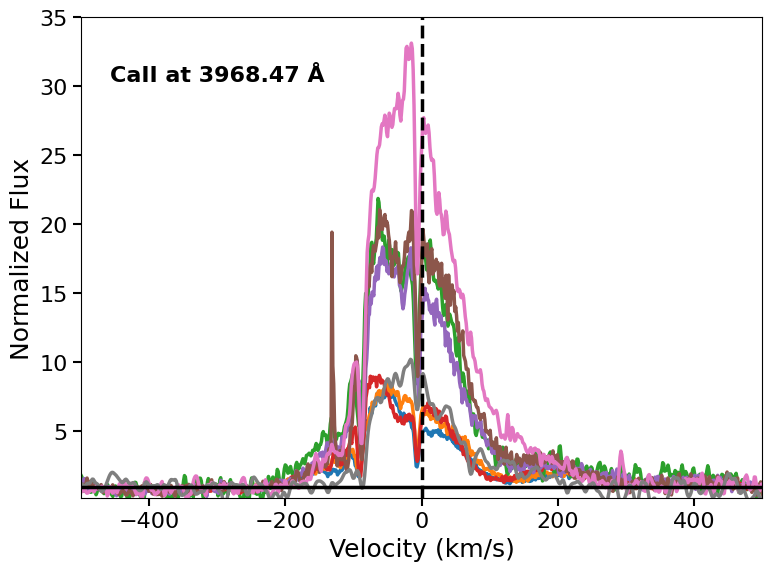}
     \includegraphics[width=0.32\linewidth]{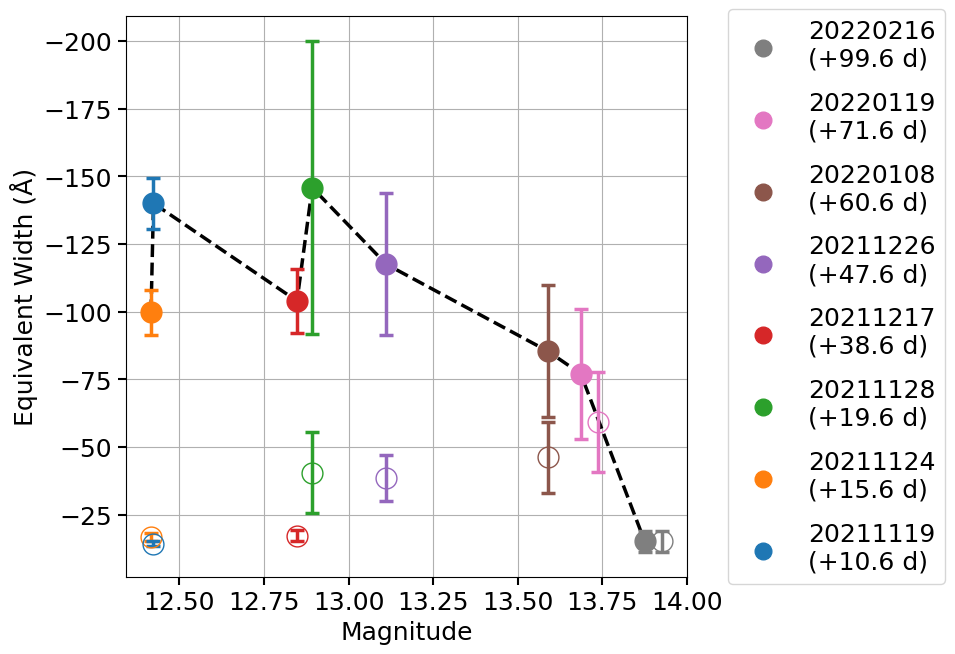} \\ \includegraphics[width=0.32\linewidth]{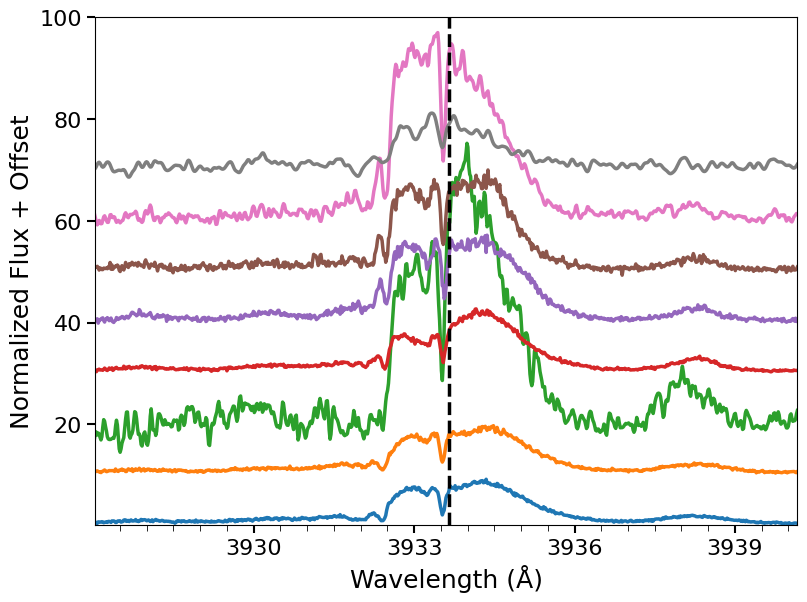}
        \includegraphics[width=0.32\linewidth]{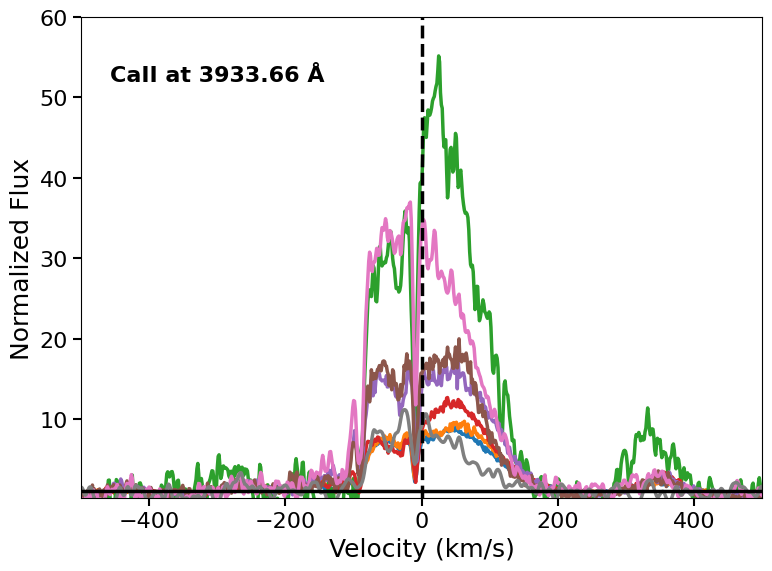}
    \includegraphics[width=0.32\linewidth]{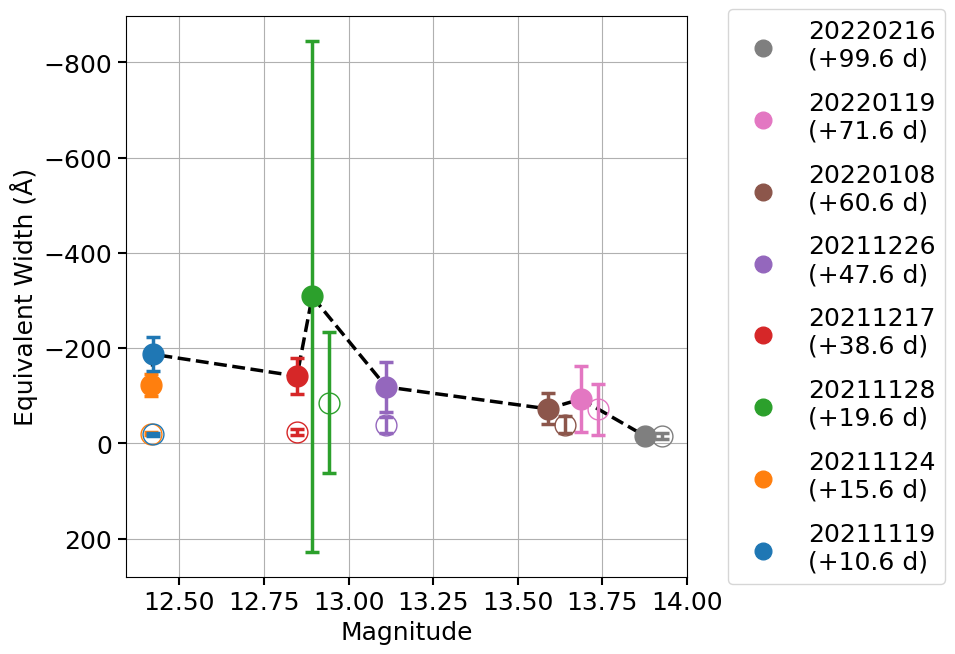}
    \caption{Spectra and equivalent width measurements of the CaII line centered at 3968.47 \AA\ (top panels) and 3933.660 \AA\ (bottom panels), {similar to Figure~\ref{hydrogen_spectra}.} As the star fades, the CaII lines become stronger in terms of line-to-continuum, except in the last spectrum.
    The {veiling-corrected} equivalent widths show {a general correlation} between source brightness and line strength. 
    }
    \label{caII-spectra}
\end{figure*}

In addition to the hydrogen Balmer lines, the \ion{Ca}{2} H\&K doublet lines are also typically strong lines in YSO spectra.
V347 Aur is no exception, and its \ion{Ca}{2} lines exhibit substantial variability over time, with profiles showing two prominent peaks whose relative strengths and velocity structure evolve as the system fades.
We note that the 3968.47 \AA\  \ion{Ca}{2} H line likely has some contamination on its red side from hydrogen $H\epsilon$ at 3970.08 \AA, which will have a morphology and strength that changes over time as discussed above for the other \ion{H}{1} lines.
The 3933.66 \AA\  \ion{Ca}{2} K profile is free of such contamination.
Unfortunately the \ion{Ca}{2} ``infrared triplet" is outside the wavelength coverage of our particular HIRES setting. 

Figure \ref{caII-spectra} shows that 
the 3968.47 \AA\ and 3933.66 \AA\ lines become {progressively} stronger in terms of line-to-continuum, and more sharply peaked as the star fades. This trend {is interrupted twice: at the 2021-11-28 epoch, the line-to-continuum strength is anomalously high relative to its neighboring epochs, particularly in the 3933.66 \AA\ line; and at the 2021-12-17 epoch, the profile briefly reverts to a wider, flatter morphology, coincident with a small photometric rebrightening. The line continues sharpening over the following three epochs before the final epoch where it returns to a broader and weaker morphology similar to that seen when the star was at its brightest.}

The velocity structure of the peaks is also notable. For the 3968.47 \AA\ line, the taller peak is blueshifted and the shorter peak is redshifted. However, for the 3933.66 \AA\ line, the taller peak is redshifted and the shorter peak is blueshifted. As the star fades further, particularly around the 2022-01-19 epoch, this pattern reverses, with the taller peak becoming blueshifted, to match the velocity structure of the 3968.47 \AA\ line. This behavior is similar to that of H$\delta$ and H$\gamma$, {where the velocity structure switches from being primarily redshifted to blueshifted.}

We also call attention to two narrow absorption lines against the broad peaks of the \ion {Ca}{2} lines, that increase in strength as the star fades.  In the earliest spectrum, the more prominent of the two features is located at -6.6 km/s with a FWHM of 6.8 km/s in the 3968.47 \AA\ line, and at -9.7 km/s with a FWHM of 8 km/s in the 3933.66 \AA\ line. The absorption feature does not show significant signs of broadening, and the velocity of the line stays constant throughout its evolution. 
The second, higher velocity absorption component is at -28.8 km/s with a FWHM of 6.4 km/s for the 3968.47 \AA\ line, and at -31.9 km/s with a FWHM of 4.7 km/s for the 3933.66 \AA\ line. In contrast to the lower velocity absorption feature, this feature shifts slightly redward as the star fades. {The narrow absorption components generally deepen as the star fades, with some exceptions.} 

{Figure \ref{caII-spectra} shows the measured and veiling corrected equivalent widths of both \ion{Ca}{2} lines (right panels). As with the Balmer series, these lines show an overall decline in veiling-corrected emission strength as the source fades, consistent with a decrease in accretion-related veiling even though the line-to-continuum has increased (center and right panels).}

\subsubsection{{Helium Lines}}

\begin{figure*}
\centering
    \includegraphics[width=0.32\linewidth]{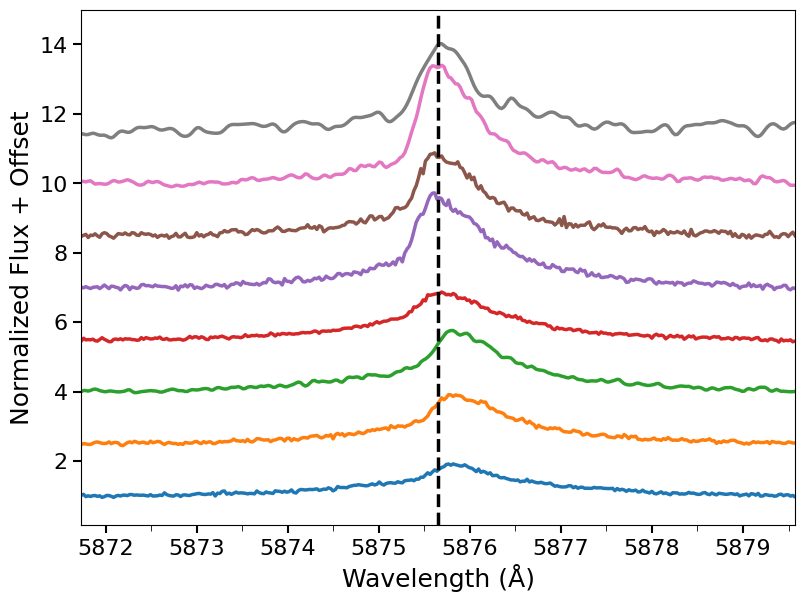}
    \includegraphics[width=0.32\linewidth]{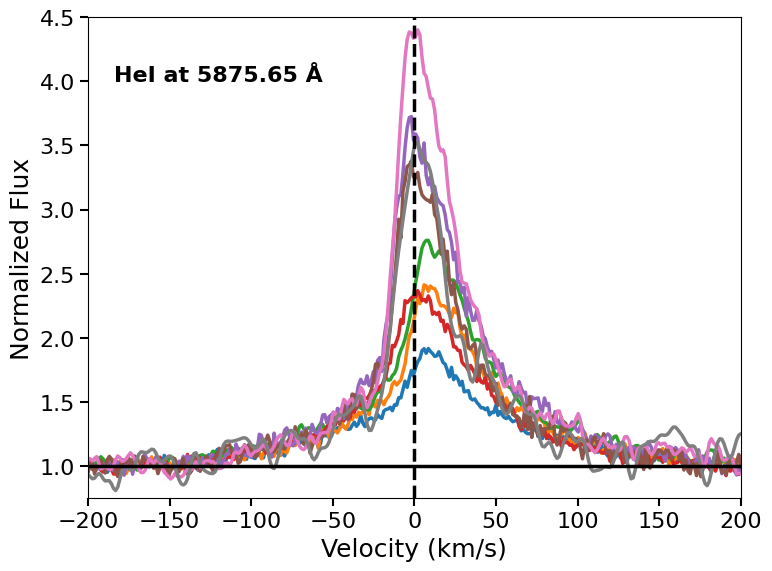}
    \includegraphics[width=0.32\linewidth]{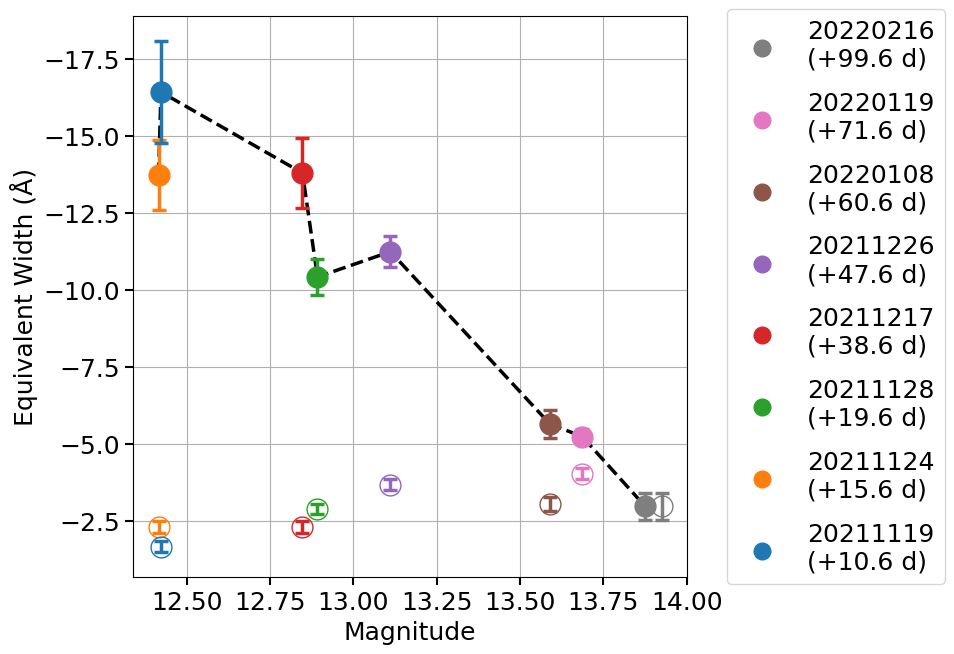}
    \\
    \includegraphics[width=0.32\linewidth]{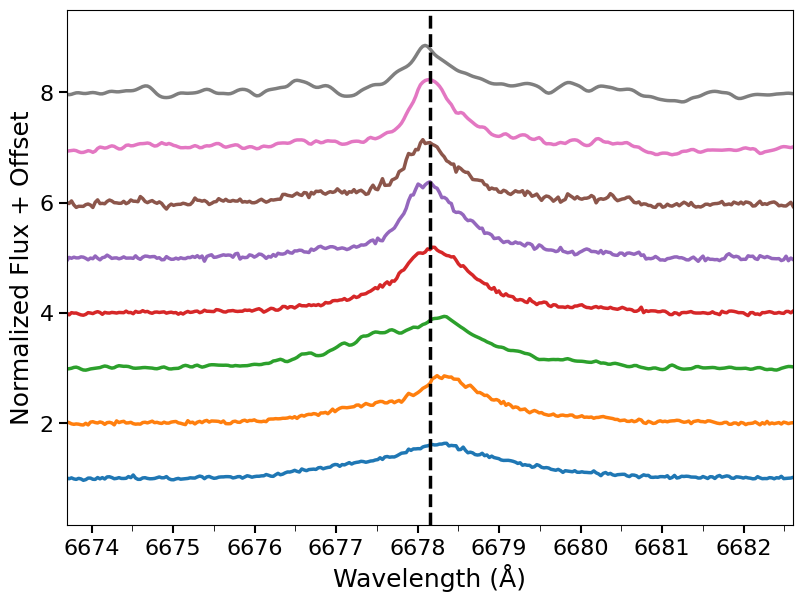}
    \includegraphics[width=0.32\linewidth]{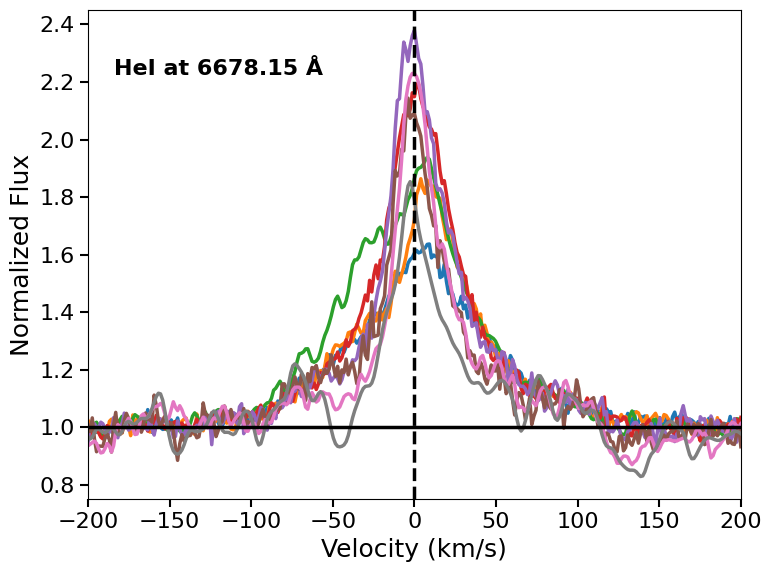}
    \includegraphics[width=0.32\linewidth]{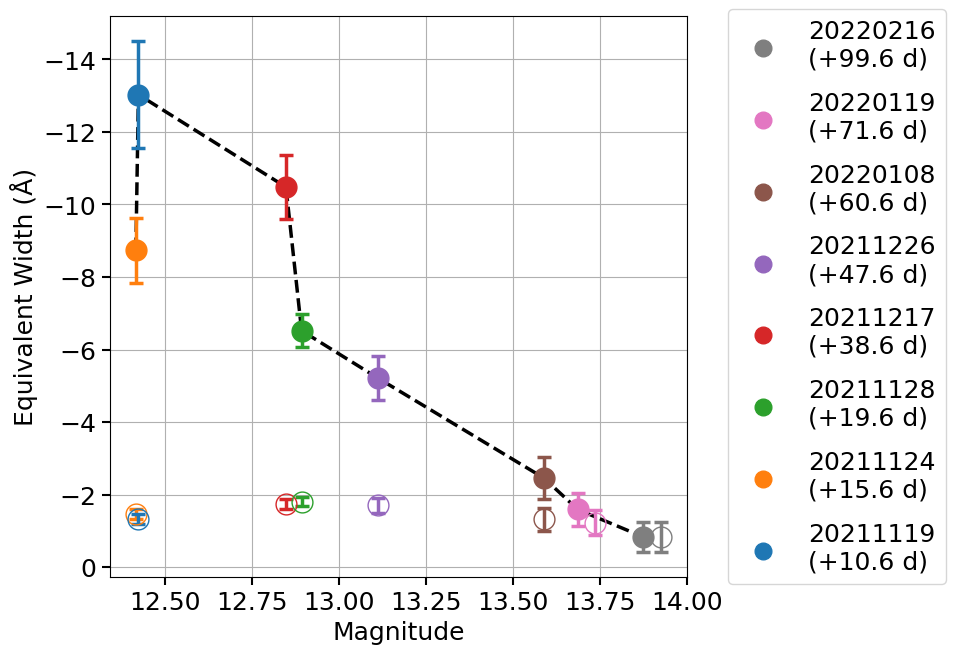}
    \caption{Spectra {and equivalent widths} for the  He I 5875.65 \AA\ line (top) and the 6678.15 \AA\ line (bottom), {similar to Figure~\ref{hydrogen_spectra}.} In general, as the star fades the He I 5875.65 \AA\ line becomes stronger relative to the continuum, and more sharply peaked. The HeI 6678.15 \AA\ profile initially follows this behavior, but then likely becomes influenced by the FeI that contributes at this same wavelength. The veiling-corrected line strengths decrease over time as the source fades. 
    }
    \label{heI-spectra}
\end{figure*}

Helium emission is not as ubiquitous as hydrogen and ionized calcium emission in young stars, but is typically present for strong accretors.
As V347 Aur fades, the \ion{He}{1} lines at 5875.65 \AA\ and 6678.15 \AA\ in our spectra exhibit subtle changes in shape and velocity. 
The line strengths change more significantly.

\ion{He}{1} is typically decomposed into two parts: the broad component (BC), which describes the base of the line, and the narrow component (NC), which describes the peak near line center \citep{beristain2001}. As shown in Figure \ref{heI-spectra}, the shape and strength of the observed BC's of the 5875.65 \AA\ and 6678.15 \AA\ lines do not change between epochs. However, the observed NC of both lines does change across epochs, driving the equivalent width changes presented in Figure \ref{heI-spectra}. 

The 5875.65 \AA\ line roughly follows the same trends as the Balmer lines, with the lines becoming stronger in terms of line-to-continuum as the star fades (with some exceptions in the {2021-12-17}, 2022-01-08, and 2022-02-16 spectra). The {line-to-continuum strength of the} 6678.15 \AA\ line {also increases as the star fades, with the exception of the final three {spectra}}. We note that the 6678.15 \AA\ \ion{He}{1} line is coincident with an \ion{Fe}{1} line, which may influence the behavior at later epochs.
{The veiling-corrected \ion{He}{1} line strengths (right panels) generally decrease in strength as the star fades, similar to the \ion{H}{1} Balmer and \ion{Ca}{2} lines.}

In terms of shape, for both lines, as time progresses, rather than strictly following the star’s brightness, the line peak shifts from being slightly redshifted to slightly blueshifted, transitioning in the {2021-12-17} spectrum. In the final spectrum, the peak of the 5875.65 \AA\ line shifts back toward the center, appearing slightly redshifted again. For the 6678.15 \AA\ line, the peak stays slightly blueshifted. 

\subsubsection{{Iron Lines}}

\begin{figure*}
\centering
    \includegraphics[width=0.32\linewidth]{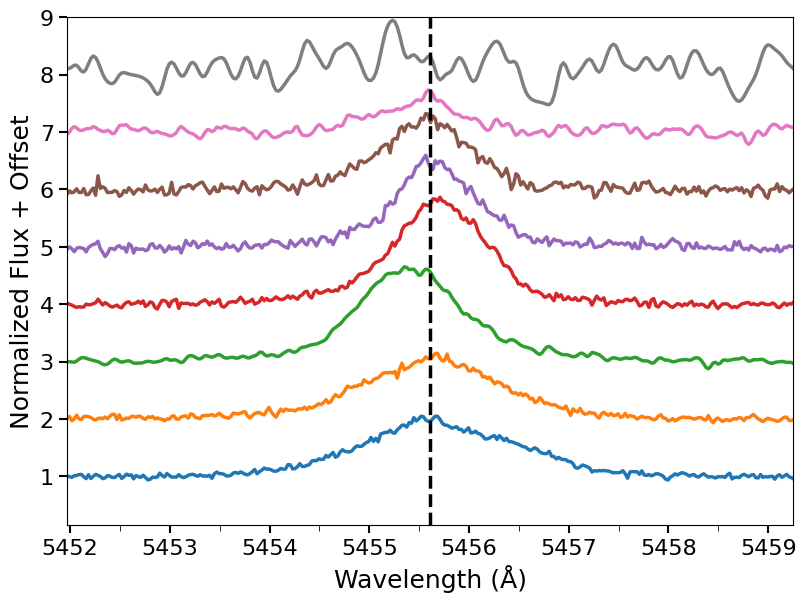}
    \includegraphics[width=0.32\linewidth]{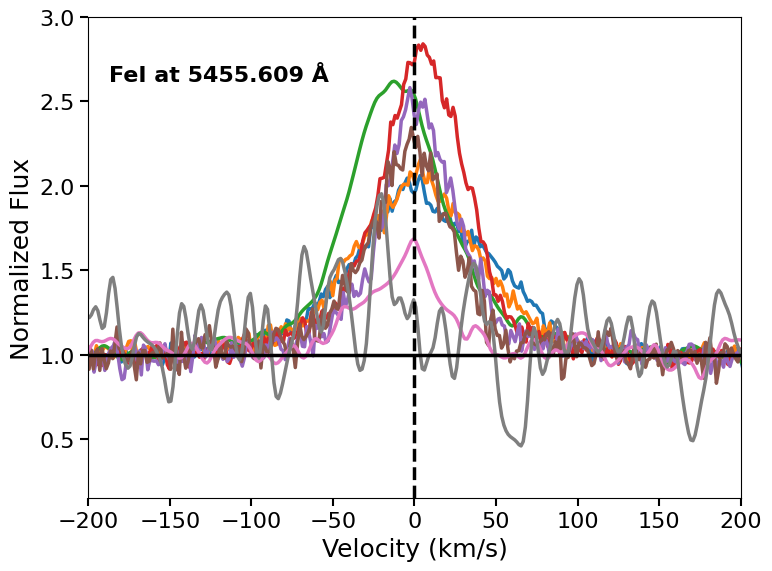}
    \includegraphics[width=0.32\linewidth]{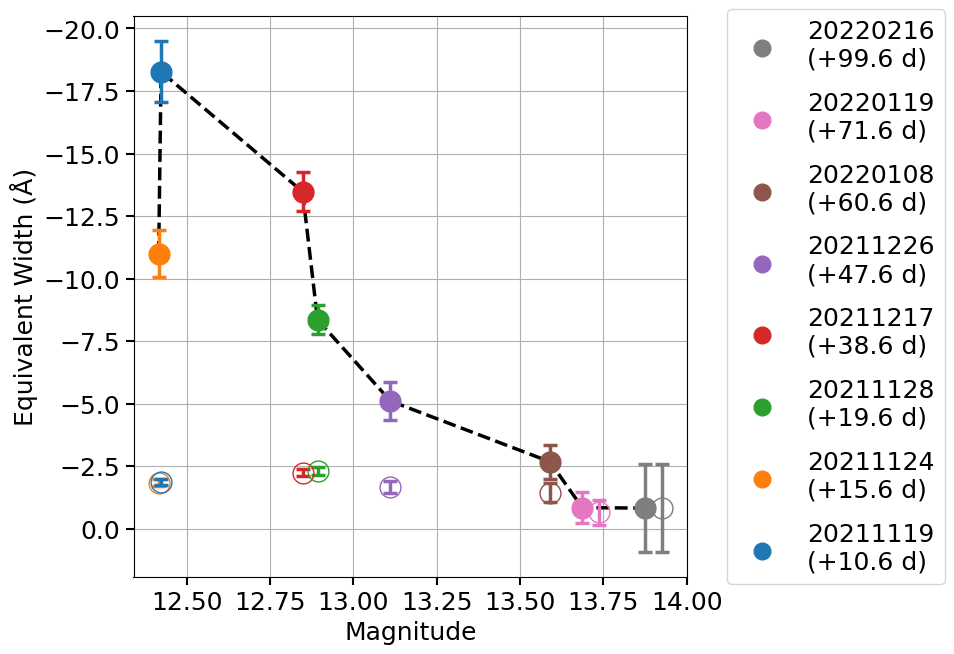}
    \\
    \includegraphics[width=0.32\linewidth]{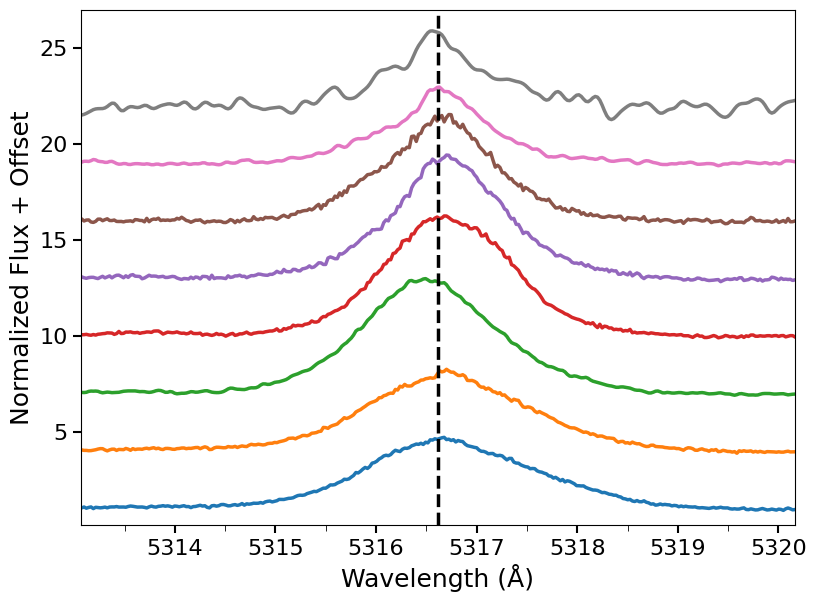}
    \includegraphics[width=0.32\linewidth]{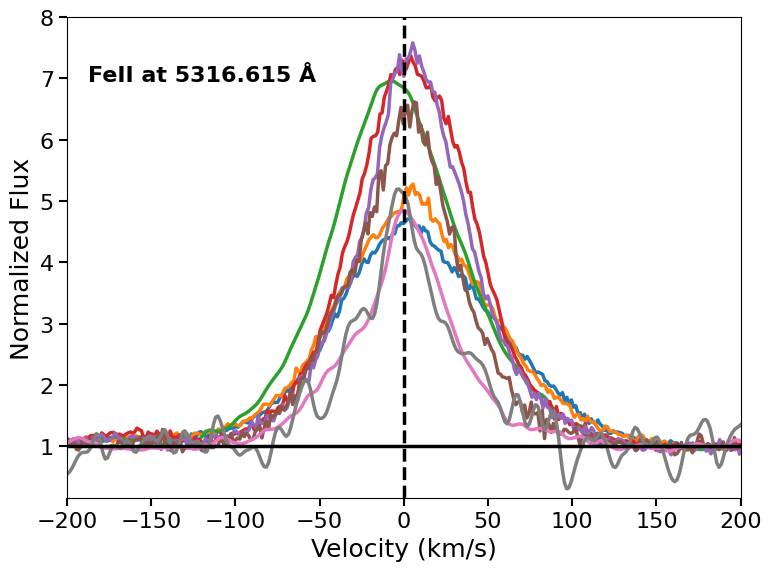}
    \includegraphics[width=0.32\linewidth]{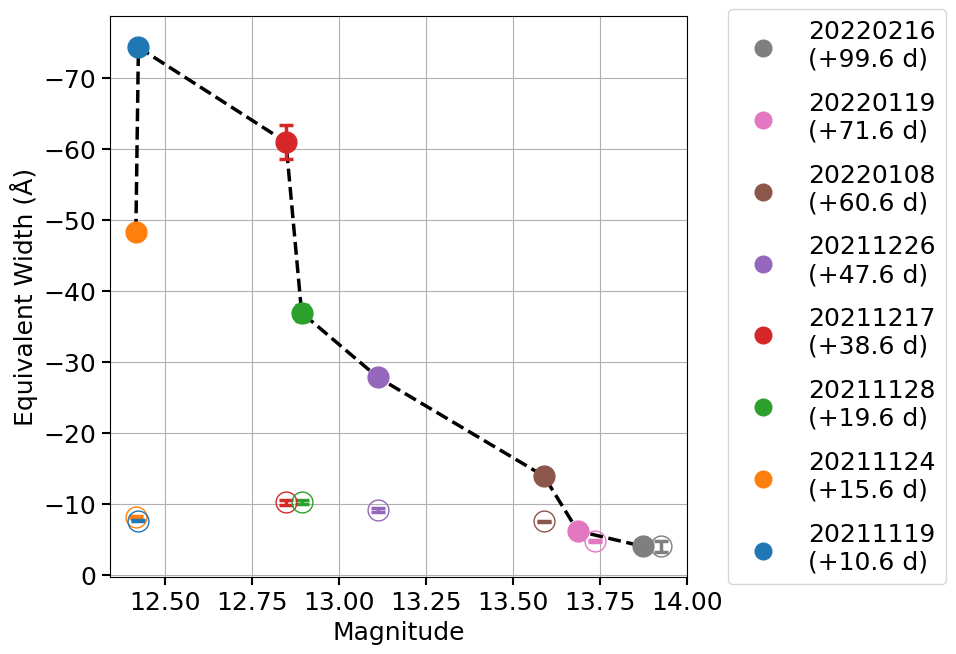}
    \caption{Spectra  {and equivalent widths} for the FeI line at 5455.61 \AA\ (top) and the FeII line at 5316.62 \AA\ (bottom), {similar to Figure~\ref{hydrogen_spectra}.}  Both lines decrease in terms of {both} line-to-continuum and veiling-corrected line strength as the star fades. We note the blueshifted asymmetry in the 2021-11-28 spectrum, which is not seen in the other line species. 
    }
    \label{FeI-II-spectra}
\end{figure*}

 V347 Aur has a rich \text{metal line} emission spectrum (Appendix~\ref{sec:lines}),
with many of the lines identifiable as \ion{Fe}{1} or \ion{Fe}{2}. We have chosen one of each as representative. 
{The equivalent widths presented in} Figure \ref{FeI-II-spectra} illustrate the declining strength of both \ion{Fe}{1} and \ion{Fe}{2} emission, with the former disappearing almost entirely while the latter persists in emission {throughout} the photometric fade and into quiescent brightness.
In the last few observations, the line widths narrow by $\sim50\%$, however, the line bases still extend to $\pm 100$ km/s, similar to the \ion{He}{1} line widths. As shown in Figure \ref{FeI-II-spectra}, the 2021-11-28 spectra is notably blueshifted for both FeI at 5455.61 \AA\ and FeII at 5316.62 \AA\, which precedes a local photometric maximum. 

\subsection{{Outflow Diagnostics}}
\subsubsection{Sodium and Potassium Lines}

\begin{figure*}
\centering
    \includegraphics[width=0.33\linewidth]{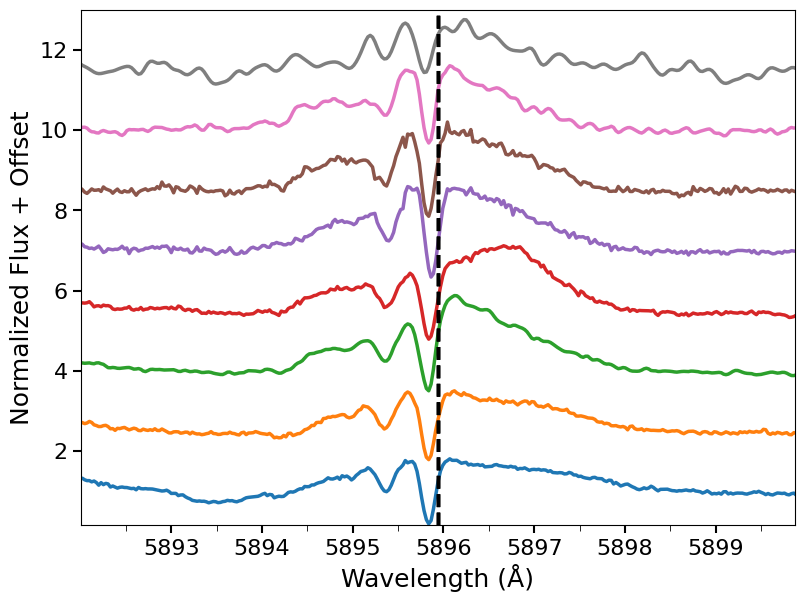}
        \includegraphics[width=0.33\linewidth]{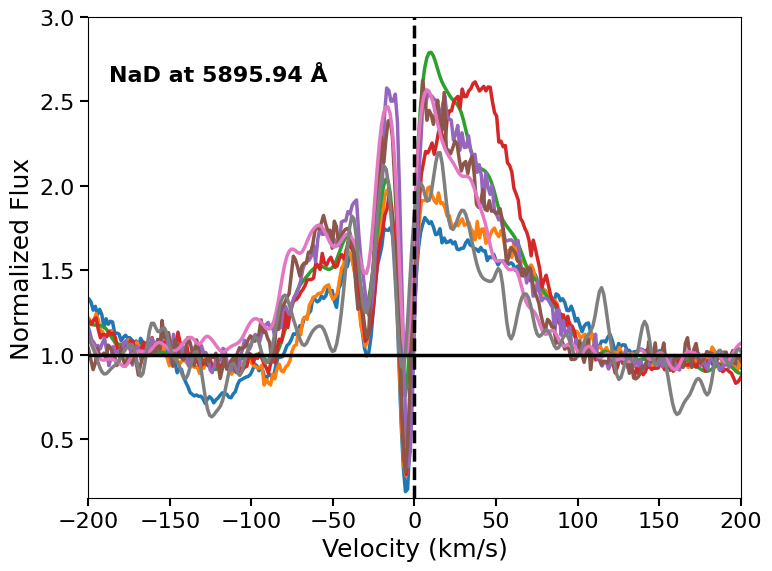}
    \includegraphics[width=0.33\linewidth]{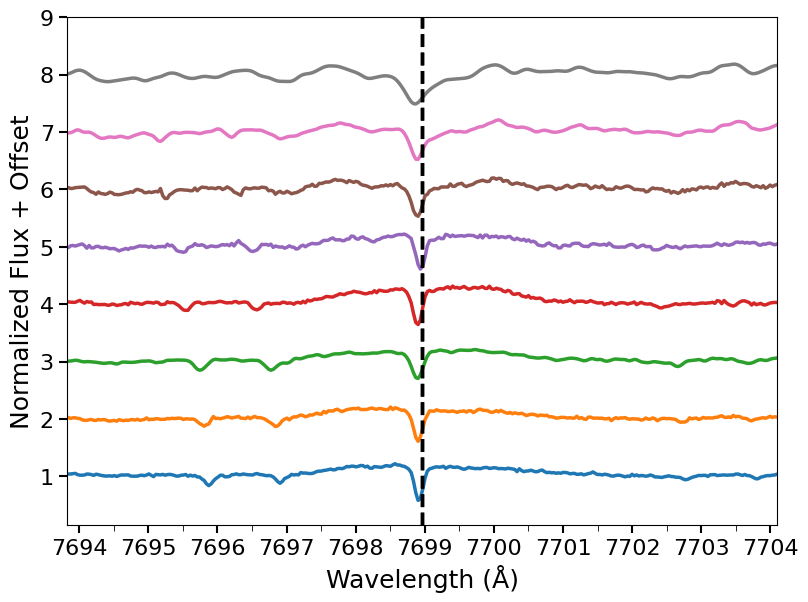}
        \includegraphics[width=0.33\linewidth]{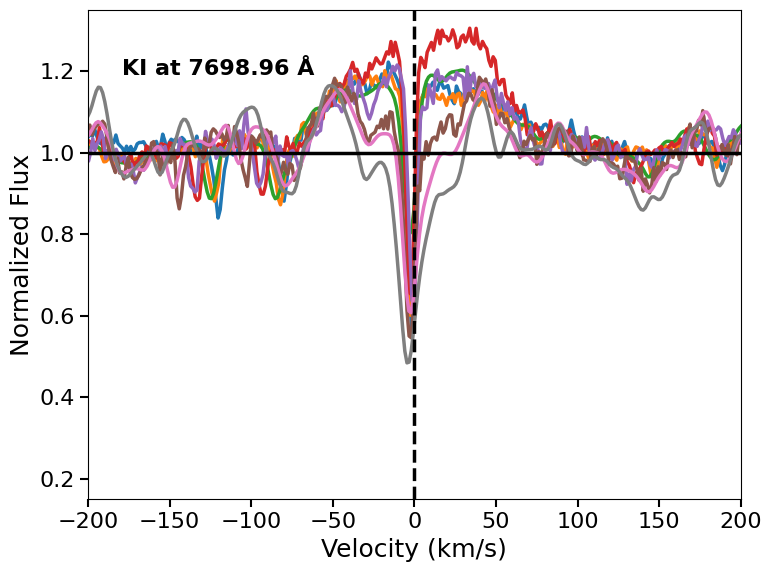}
    \caption{Spectra for the NaD line centered at 5895.94 \AA\ (top) and KI line centered at 7698.96 \AA\ (bottom). The narrow absorption features around -29 and -5 km/s in the NaI D line, and the higher velocity (-29 km/s) narrow feature in the KI line maintain stable velocities across all epochs. The lower velocity absorption feature in KI shows a shift to slightly higher velocities (-2 to -4 km/s) as the star fades.
 }
    \label{wind_lines_spectra}
\end{figure*}

The NaD and the KI doublet lines, while also tracing accretion \citep{muzerolle2001} are frequently associated with winds in YSO spectra \citep{edwards1994}. These features are illustrated in Figure \ref{wind_lines_spectra}.

Both the NaI D line centered at 5895.94 \AA\ and the KI line centered at 7698.96 \AA\ exhibit narrow blueshifted absorption components seen against the broader emission spectrum. 
The morphology is  similar to that observed in the CaII lines. 
These absorption features have a {deep} low-velocity component and a shallower high velocity component. 
Both components are relatively stable in velocity across all epochs, but the absorption depths vary. 
The morphology of the line wings also varies, suggesting temporal changes in the structure of the accretion/wind. 
In the NaI D line, the lower velocity absorption component is at -5.3 km/s with a FWHM of 9.7 km/s measured from the first spectrum collected. The higher velocity component is at -29.1 km/s and is broader than the low-velocity component, with a FWHM of 19.3 km/s measured from the first spectrum collected. 
We measure no significant velocity shifts {or broadening} in the NaI D narrow absorption.

We also note that the narrow components in NaI D are absorbing against a broader emission component that is asymmetric, peaking on the red side. 
In the KI line, by contrast, the narrow absorption components are seen against a broader emission peak that appears symmetric between the red and blue sides. 
Also unlike NaI D, the morphology of the KI narrow absorption components evolves slightly over time.

In the KI line, the lower velocity narrow absorption component line is initially at -1.8 km/s, but shifts to slightly higher velocities as the star fades with -3.7 km/s measured in the final spectra. Over the final three epochs, {this lower velocity} absorption {component} deepens and broadens, from FWHM of 5.4 km/s in the earlier spectra to 9.3 km/s in the final spectrum. We interpret this low-velocity feature as transitioning from tracing only wind absorption, to also revealing the zero-velocity stellar photospheric absorption line. This behavior is similar to that of \ion{Li}{1}, where a decline in the continuum flux over time allows the underlying photospheric absorption to emerge. 
Meanwhile, the higher velocity narrow absorption component is blueshifted to -28.6 km/s across all epochs. 
This feature broadens only slightly as the star fades to a FWHM of 6.5 km/s measured from the final spectrum. 

Due to the complex structure of the lines, we did not compute equivalent width measurements. 

\subsubsection{Forbidden Lines}

Optical forbidden lines of [O I] and [S II] trace outflowing gas in young stars,
{with the lines formed in the shock interaction of accretion-driven jets 
and gas in the ambient environment}. 

Both of these lines are doublets, but we present here
only the stronger line in each case, specifically the [OI] 6300.30 \AA\ line
(6363 \AA\ exhibits the same morphology only weaker, with the expected 3:1 ratio), 
and the [SII] 6730.82 \AA\ line 
(6716 \AA\ shows the same morphology, but is weaker with 2:1 ratio).
We also note that V347 Aur has weak [CaII] 7291.47 and 7323.89 \AA\ emission, 
but we do not discuss these lines any further.

\begin{figure}
\centering
    \includegraphics[width=\linewidth]{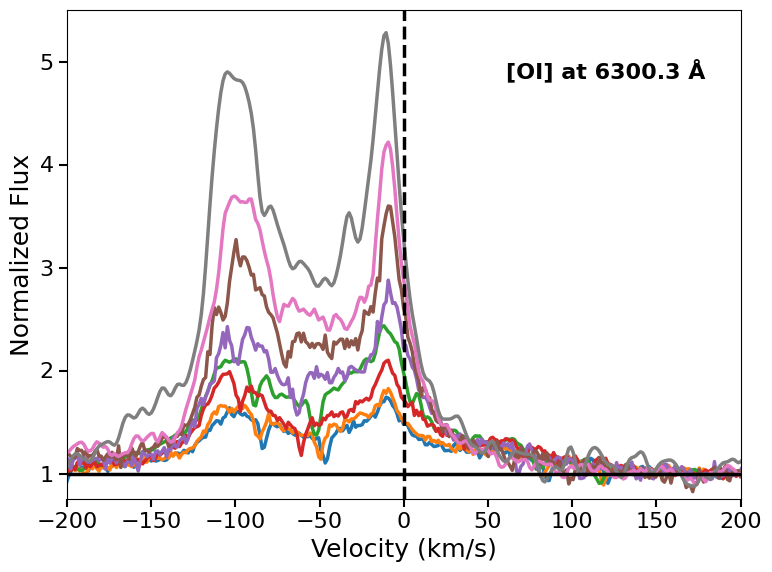}
    \includegraphics[width=\linewidth]{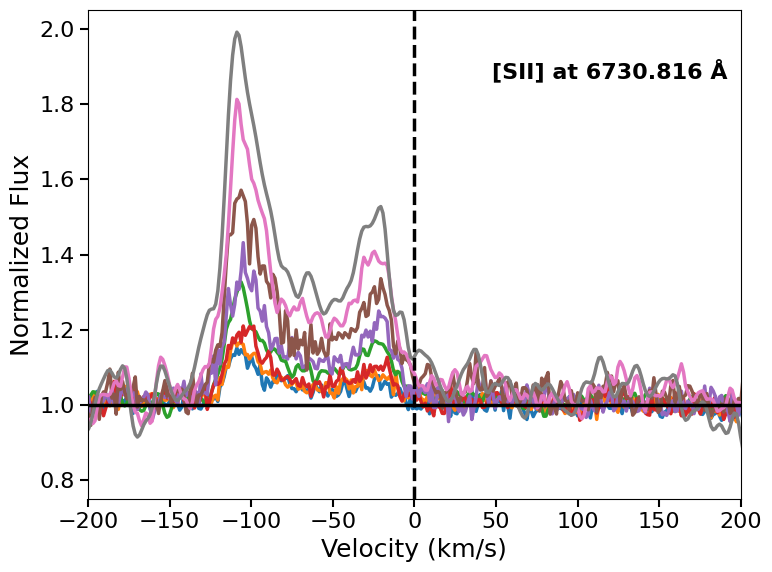}
    \caption{Velocity spectra for the [OI] 6300.30 \AA\ line (top) and the [SII] 6730.82 \AA] line (bottom). Both lines consistently exhibit blueshifted behavior. The [OI] line has broad high-velocity wings extending out to +90 km/s and -200 km/s, whereas the [SII] line does not exhibit this broad base.}
    \label{forbidden_lines_vspectra}
\end{figure}

{The time series of [OI] and [SII] profiles is} 
shown in Figure \ref{forbidden_lines_vspectra}.
Both species exhibit distinct low-velocity and high-velocity components, 
with all peaks entirely blueshifted. 
As the star fades, both peaks in both species become more sharply peaked and narrower. 

The [OI] profile has two components of comparable strength. The line-to-continuum intensity of the higher velocity component reaches a peak 0.92 times that of the lower velocity component in the 2022-02-16 (last) spectrum. The high-velocity component is centered at $-100$ km/s and the low-velocity component near $-8$ km/s. Broad high-velocity wings extend out to $+90$ km/s and $-200$ km/s.  The implied centroid for the broad part of the low-velocity component would thus be about $-55$ km/s, consistent with formation in a turbulent wind. The overall morphology of the [OI] line profile, with a broad shoulder between the low-and high-velocity peaks which have comparable line-to-continuum strength, resembles the morphology seen in the source AS 353A \citep{Banzatti2019}.  These authors also present [OI] FWHM values as a function of the [OI] centroid velocities for their sample.  
{Objects with the most blueshifted centroids tend to have the broadest high velocity component, while less blueshifted low-velocity components span a wide range of FWHM values, from comparably broad to narrower than the high-velocity component population.} This is broadly consistent with what we observe in V347 Aur, where the peak of the high velocity component is broader than the peak of the low-velocity component by a factor of 1.76 (measured from the 2022-02-16 spectrum), {though the base of the low velocity component is comparably broad to that of the high velocity component.}

The [SII] profile, by contrast, exhibits two entirely blueshifted components at $-105$ km/s and another at $-25$ km/s. The higher velocity component is always larger than the lower velocity component, with the relative line-to-continuum intensity {between the two components} peaking at 1.3 in the 2022-02-16 spectrum.  
Furthermore, there is no obvious broader element in the low-velocity component, and specifically no redshifted emission as we observe in the [OI] line. 

Due to the complex structure of the {forbidden} lines, we did not compute equivalent width measurements.

\section{Discussion}
We have analyzed a sequence of high dispersion optical spectra 
for the quasi-periodic outbursting young stellar object V347 Aur,  
taken during the decline phase of a recent brightening event (Figure \ref{light_curve}).
The multi-epoch dataset shows systematic change in spectral line strength and morphology
that can be correlated with the photometric brightness of the source.
{In this section we discuss the accretion diagnostics, then the outflow and wind diagnostics,
and finally the implications of our results.}

\subsection{{Accretion Continuum and Line Emission}}

During the earlier epochs of our spectroscopic data set, when the source was brighter, photospheric absorption features including TiO and \ion{Li}{1} are strongly veiled and nearly undetectable. These features re-emerge during quiescence, when the excess continuum emission is reduced, allowing the underlying photospheric spectrum to be revealed (Figures \ref{tioband} and \ref{liI_spectra}).
The observed trend is consistent with a scenario in which accretion is stronger during the earlier epochs, when the source is near its peak brightness, and excess continuum emission from hot gas veils the underlying photospheric spectrum, filling in absorption features.  As the accretion rate drops and the source fades, the veiling diminishes, and the intrinsic absorption features become more detectable.

Balmer line emission is a long-standing tracer of the accretion flow \citep{muzerolle2001}. As V347 Aur fades, the Balmer lines 
generally decrease in strength and the line profiles appear sharper and more structured (Figure \ref{hydrogen_spectra}).

The Balmer profiles also undergo morphological evolution. Early spectra are asymmetric towards the red side, or have double peaked profiles preferentially in the red, while later epochs show enhanced blue-side emission and the development of a more complex blueshifted emission. This may reflect early emission dominated by magnetospheric accretion and post-shock regions and later, reduced veiling allows us to see blueshifted wind emission and multiple velocity components. 

Broad \ion{Ca}{2} emission is also generally attributed to the magnetospheric accretion flow \citep{kwan2011}. 
In V347 Aur, the Ca II emission profiles display two prominent components, a redshifted and a blueshifted peak, whose relative strengths vary across epochs and between the two lines (Figure \ref{caII-spectra}).  
The Ca II 3968.47 \AA\ line consistently exhibits a stronger blueshifted component than redshifted component. In contrast, the Ca II 3933.66 \AA\ line shows the opposite behavior at early epochs, with the redshifted component dominating, until the 2022-01-19 epoch, when this trend reverses.
Although these lines share a lower level, the upper level of the K (3933.66 \AA) line is slightly higher than the upper level of the H (3968.47 \AA) line, the latter of which can have contamination from H$\epsilon$ at 3971.20 \AA.
The alternating dominance between redshifted and blueshifted emission components is {suggestive of} a magnetospheric accretion geometry where the emitting region co-rotates with the star. As the system rotates, the orientation of the accretion funnels and associated emitting regions changes relative to the observer, leading to variations in the line of sight velocities. At times, emission from material moving toward the observer becomes more prominent, producing stronger blueshifted components, while at others, emission from receding material dominates, enhancing the redshifted side of the profile. In addition, accretion flows are not perfectly smooth. Localized high density regions or clumping within the funnel flows could further contribute to the relative contributions of the redshifted and blueshifted components, contributing to the observed asymmetries. {However, our observations are sparse relative to the $\sim 4.1$ day rotation period \citep{donati2024}; thus we cannot confirm this interpretation.}

\ion{He}{1} emission \citep{edwards1994,JohnsKrull1997} 
is formed both in the accretion funnel flows {and in the outflowing wind.} The accretion shock forms the narrow component (NC)
in the post-shock gas at the footprint of the accretion column, 
and the outflowing wind contributes the wide range of velocities in the broad component (BC), as described in \cite{beristain2001,kwan2011}. 
In V347 Aur, the \ion{He}{1} {veiling-corrected} equivalent widths {decrease as the star fades (Figure \ref{heI-spectra}).} Additionally, the lines do not display similar behavior between their NC and BC. The BC is traditionally found to vary in sync with other accretion tracers, such as the veiling \citep{bouvier2007} or the Balmer lines \citep{Zsidi2025}; however, we do not see this trend in our spectra. The NC has been found to show notable amplitude variations \citep{Zsidi2025}; this is consistent with our spectra.

We infer that as the star fades, the \ion{He}{1} emission becomes less dominated by accretion and increasingly influenced by blueshifted material in the wind, shifting the line peaks towards the blue.

\subsection{{Wind and Outflow Diagnostics}}

{In H$\alpha$}, there is blueshifted absorption superimposed on the broad emission {profile, seen} at different velocities over the different spectral epochs. 
{We call attention to the feature at -40 km/s in the 2022-01-19 epoch,
and note that similar features are apparent at other epochs.}
This morphology {could be} a sign of mass outflow in the form of a wind {\citep{JohnsKrull1997}. 
\cite{hartmann1994} suggest that the blueshifted absorption in Balmer lines arises in a low-density, optically thin wind, which they speculate originates from outside the magnetosphere, whereas the majority of the line emission originates in the infalling flow. This is consistent with wind models from \citep{Kurosawa2012}, where blueshifted absorption components are caused by the conical-shell wind.}

$H\alpha$ has a higher optical depth than $H\gamma$ and $H\delta$, which may explain why the higher Balmer lines retain clearer signatures of the underlying emission structure. 
\citet{calvet1992} produced central and low-velocity narrow absorption against broad emission profiles for Balmer lines in magnetospheric infall/accretion models,  when they are viewed at certain viewing geometries i.e. inclinations. 
{Here we note that} \cite{donati2024} report inclination $i=26\pm5$ deg for V347 Aur 
based on their measured projected rotational velocity and an inferred stellar radius. 
{However, the features we see occur over a range of velocities, that are different at different epochs.
It is also possible that the variable location of the narrow absorption
is a consequence of the complex line radiative transfer in a rotating, accreting magnetosphere.}

{Besides the narrow blueshifted features in H$\alpha$ that are time-variable,}
V347 Aur also exhibits persistent, narrow, blueshifted absorption lines that are seen 
against the broad emission profiles in several other species: 
CaII, NaI D, and KI (Figures \ref{caII-spectra} and \ref{wind_lines_spectra}).  
This also nominally points to self-absorption by cooler gas within the winds.

It should be noted that there is interstellar contribution to the absorption in these Ca II, NaI D, and KI lines.
NaI D and KI column densities are usually well-correlated \citep[e.g.][]{Welty2001} 
with CaII tracing higher temperature gas along the line of sight. 
For a sample of YSOs in Taurus, \citet{pascucci2015} argued  that narrow  {\it low-velocity}
absorptions in NaI D and KI are neither circumstellar nor from local (to the sun) interstellar matter, 
but rather are formed in the cooler larger-scale Taurus molecular cloud.
In V347 Aur, the narrow absorption components 
have observed FWHM values ranging up to 19.3 km/s. 
These widths can be compared to values observed in the local ISM, which out to 300 pc 
do not exceed FWHM=5.8 km/s ($b=3.5$ km/s)
as reported by \cite{welsh2010}. {The relatively broad width of these narrow absorption components indicates that they cannot solely be attributed to the interstellar medium, supporting the self-absorption interpretation.}

The narrow absorption components are largely stable in velocity over time, indicating {a steadier wind geometry with the absorbers likely tracing gas further from the star, where velocities have reached a roughly constant value.} In CaII, two persistent absorption components are present for both the H and K transitions. The low-velocity components are at about -8 km/s and the high-velocity components are at about -30.5 km/s.
NaI D likewise exhibits two narrow absorption components, at -5.3 km/s and -29.1 km/s, 
which are similar velocities to the CaII H and K lines. 
For both CaII and NaI D, these velocities are fixed across epochs. 
In K I, the lower velocity component shifts to slightly higher velocities (from -1.8 to -3.7 km/s) as the star fades,
but the higher velocity component at -28.6 km/s is similarly stable. 
With the exception of the low-velocity KI feature, the consistent absorption velocities across epochs favor a steadier wind geometry, rather than a ``dynamic" wind. 

While the absorption components show some correspondence across species -- particularly the component near -29 km/s, their velocities are not identical between line species, particularly the lower velocity features. This behavior implies different formation locations in the wind. In this scenario, CaII, with an ionization potential of 11.9 eV, could persist under denser conditions closer to the wind base, while NaI and KI, with lower ionization potentials (5.14 eV and 4.34 eV respectively), could trace cooler material further out as they cannot survive as neutrals close to the host star. Additionally, the CaII absorption lines are at higher velocities than NaI D and KI, so may be tracing the wind closer to the star than further out, when it has slowed down. 

The lowest-velocity KI absorption component shifts slightly towards higher velocities, deepens, and broadens, suggesting a transition from tracing wind absorption to revealing the underlying stellar photospheric line as continuum veiling declines. This is similar to the behavior of the \ion{Li}{1} line, where a decline in continuum veiling over time allows the underlying photospheric absorption to emerge. Thus, not all low-velocity absorption necessarily originates in the outflow. 

The existence of multiple narrow absorption components in these species is reminiscent of RU Lup, which displays narrow absorption components in the broad emission lines of CaII H and K and NaI D {\citep{armeni2025}}. Similar to V347 Aur, the intensities and line profiles of these lines stay relatively constant, independent of the brightness variations of the star. \cite{Gahm2013} conclude that this behavior in RU Lup indicates that the gas flow must originate far out from the star and enclose the stellar wind. Additionally, they concluded that the blueshifted absorption spectrum could be related to irregularities such as pockets or shells of higher density in the disk wind. This is consistent with our observations of multiple narrow absorption components that all reside at similar velocities, yet differ slightly between lines. 

The [O I] and [S II] forbidden lines are widely used diagnostics of outflowing gas 
in young stellar objects  \citep{edwards1987,Frank2014}. 
High-velocity components are generally interpreted as originating in a fast outflow or jet,
while low-velocity components are attributed to an MHD wind.  
In V347 Aur, these lines are consistently blueshifted (Figure \ref{forbidden_lines_vspectra}),
consistent with the outflow {geometry in which the receding (redshifted) component is} at least partially obscured by the disk. 

There is no noticeable change in the locations of the peaks, indicating that the jet persists at a similar velocity, while the observed line strengths relative to continuum 
are affected by changes in the accretion rate. 
The {persistent} line morphology suggests a stable outflow geometry. However 
as the star fades, both peaks in both species become narrower and more sharply peaked, 
suggesting a transition to a less turbulent wind. 

The [OI] line exhibits broad high-velocity wings {on both the redshifted and blueshifted sides}. 
In contrast, the [SII] line lacks a redshifted emission wing. 
The differing behavior between [OI] and [SII] can be interpreted in terms of line critical densities, and thus the regions of the outflow they trace. 
The [OI] line has a higher critical density than [SII], thus the [OI] transition is able to occur in dense gas that is closer to the star, including the base of the jet and inner disk wind. In these regions, some emission from the receding side of the flow may remain visible, producing the observed redshifted wing. The [SII] transition has a lower critical density, thus it is not able to occur in regions as dense as the [OI] transition, and must occur further out along the outflow, where the receding side is more effectively hidden by the disk or other circumstellar material. As a result, the redshifted emission may be suppressed in [SII] -- as we observe.

Our observations are consistent with previous work demonstrating that [OI] traces both a dense inner wind and the base of a jet, while [SII] probes a lower density and more extended portion of the jet, with the disk geometry such that the redshifted component of the jet is blocked. 

\subsection{{The Binary-Induced Pulsed Accretion Model for V347 Aur}}

{As
discussed in detail by  \citet{dahm2020} who proposed it, and by \citet{donati2024} who confirmed it, 
a viable model for the photometric and spectroscopic behavior of V347 Aur is an orbiting binary 
system in which a burst of accretion is dynamically induced near the periastron passage
\citep{artymowicz1996,munoz2020}.  
Other well-studied examples of this phenomenon are: DQ Tau, UZ Tau E, and TWA 3.
In these systems, the enhanced accretion is assumed to occur onto the primary.
We note that our observations of V347 Aur do not spatially or spectrally resolve the binary.
}

{\cite{donati2024} interpret their data in terms of the binary influencing the magnetic field. They speculate that pulsed accretion, driven by the companion's periastron passages, enhances shearing in the convective layers of the star and amplifies the large-scale toroidal field, a process they describe as a ``pulsed dynamo." They also discuss that the system may host an inner disk cavity that could pump the observed orbital eccentricity of the companion, through interactions with the disk. They further propose that this eccentric orbit would, in turn, carve a deeper cavity over time. This process could explain the enhanced accretion episodes observed at periastron passage in both the photometric light curve and the corresponding signatures in the spectra.}

{Our measurements are of spectral line changes in V347 Aur as it recovers 
in brightness from a photometric outburst that is attributed to a
temporarily increased accretion rate. The observations cover a time period of months 
during which the accretion rate declines by over an order of magnitude, 
and are the first to link the photometric and emission line behavior in this source.
Our results can inform models of pulsed accretion in binary systems.  
They are also applicable to the more general area of accretion variability in young stellar objects.
}

\section{Summary} 

We have presented the results of time series optical spectroscopy at high dispersion
obtained during the decay phase of an accretion burst in the young stellar object V347 Aur. 

Our spectral dataset consists of eight epochs.
We analyze the lines of \ion{Li}{1} (6707.86 {\AA}), \ion{H}{1} (H$\alpha$, H$\delta$, H$\gamma$), HeI (5875.62 {\AA}, 6678.15 {\AA}), CaII (3968.47 {\AA}, 3933.66 {\AA}), FeI (5455.61 {\AA}), FeII (5316.62 {\AA}), KI (7698.96 {\AA}), NaD 5895.94 {\AA}), [OI] (6300.30 {\AA}), and [SII] (6730.82 {\AA}). We have measured line center shifts, line broadening, equivalent widths, and detailed the changes in line profile morphology.

As V347 Aur fades, photospheric absorption features deepen, 
consistent with a reduction in accretion driven continuum veiling. 
{The broad emission lines all generally increase in their observed line-to-continuum strength as the source fades, but decrease in veiling-corrected equivalent width with the presumed decreasing accretion rate. The Balmer emission lines transition to a more blueshifted profile as the star fades, as do the HeI lines.} The CaII line profiles show alternating dominance between redshifted and blueshifted components, which we interpret as the rotation of the system and changing viewing angles of the magnetospheric funnel flows.
We also detect narrow, blueshifted absorption components in CaII, NaD, and KI
that exhibit correlation with one another. 
Forbidden emission lines have multiple components and are consistently blueshifted and trace the kinematics of outflowing material. The [OI] emission likely originates from a denser, inner wind region, while the [SII] line probes lower density, more extended jet material. The absence of strong redshifted emission, particularly in [SII], is consistent with a disk geometry that partially obscures the receding side of the outflow. 

{In this paper we have mapped the changes in observed accretion and outflow diagnostics
in a periodically outbursting YSO that experiences order-of-magnitude variations
in its accretion rate on time scales of months.
The V347 Aur system is a useful proxy both for the ubiquitous short-term accretion variability observed in YSOs, and for their long-term secular evolution.}

\begin{acknowledgements} 
We thank Suzan Edwards for useful comments that helped us improve our presentation,
{and our referees for their attention and corrections} to the manuscript.
\end{acknowledgements} 

\renewcommand{\refname}{References}
\bibliographystyle{aasjournal}
\bibliography{references}

\appendix

\section{Inventory of Emission Lines in V347 Aur}\label{sec:lines}

\begin{figure}
\centering
    \includegraphics[width=0.55\linewidth]{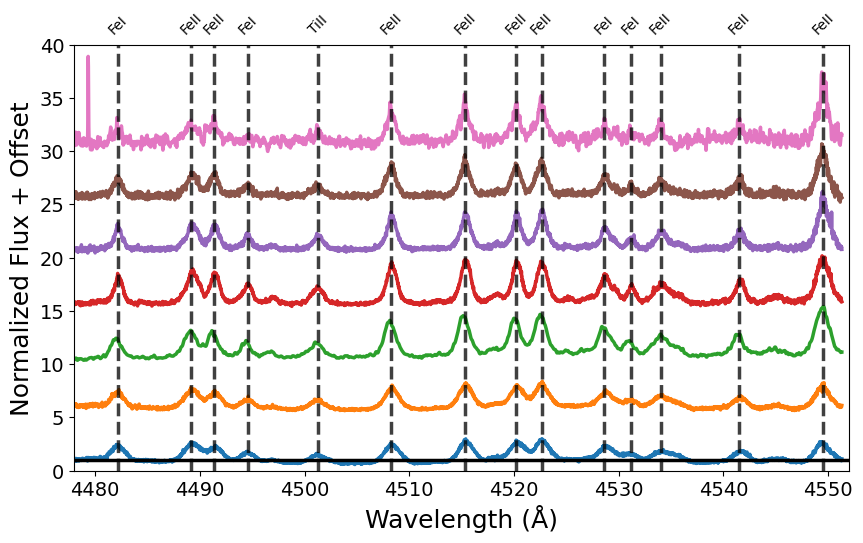}
    \caption{Time series spectra for one HIRES order showing numerous emission lines including from FeI, FeII, and TiII.}
    \label{fig:inventory}
\end{figure}

Figure~\ref{fig:inventory} illustrates the general emission-line nature of V347 Aur 
during its outburst (bottom) to more quiescent (top) stages.
Table~\ref{tab:species} provides a full inventory of the identified emission lines.

\input{table}

\end{document}

%% file: table.tex
\setlength\LTleft{\fill}
\setlength\LTright{\fill}
\setcounter{LTchunksize}{600}

\startlongtable
\begin{longtable}{c|c}
\caption{Detected Emission Lines} \label{tab:species} \\
\hline
\textbf{Species} & \textbf{Wavelength [\AA]} \\
\hline
\endfirsthead

\hline
\textbf{Species} & \textbf{Wavelength [\AA]} \\
\hline
\endhead

\endfoot
\hline
\endlastfoot

H-19       &           3686.83 \\   
H-18       &            3691.55 \\     
H-17        &           3697.16 \\     
H-16       &            3703.86\\     
TiII:     &            3706.23 \\      
H-15       &           3711.98 \\       
FeI & 3719.95 \\ 
H-14       &           3721.94 \\     
FeI       &            3727.03 \\      
H-13  &  3734.37\\ 
H-12 & 3750.15 \\
FeI        &            3758.24 \\  
FeI & 3763.7888 \\
H-11       &            3770.63 \\  
NiI        &            3783.53 \\   
FeI        &            3787.88 \\ 
H-10       &            3797.90 \\  
NiI        &            3807.14 \\  
MgI        &            3832.31 \\
H-9        &            3835.39 \\
MgI        &            3838.30 \\
FeI        &            3841.05 \\
FeI        &            3852.57 \\
FeI        &            3859.92 \\ 
FeI  & 3872.5 \\
H-8        &            3889.05 \\  
FeI        &            3895.64 \\   
FeI        &            3900.53 \\ 
FeI        &            3902.94 \\  
FeII       &            3905.63 \\  
TiII       &            3913.47 \\   
FeII & 3914.12 \\
FeI        &            3917.18 \\   
FeI        &            3920.26 \\  
FeI        &            3922.91 \\  
FeI        &            3927.92 \\  
FeI        &            3930.29 \\  
CaII       &            3933.66 \\  
FeII       &            3938.34 \\  
AII    &           3944.01 \\ 
FeI         &            3945.12 \\   
HeI        &            3964.71 \\
CaII       &            3968.47 \\
H$\epsilon$ &            3970.84 \\
FeII       &            3974.17 \\
FeI        &            3977.74 \\
MnI        &            3982.16 \\ 
FeI        &            3983.96 \\
FeI        &            3995.20 \\
FeI        &            3997.40 \\
FeI        &            4001.66 \\
FeI        &            4005.24 \\
FeI        &            4009.72 \\
CrII       &            4012.38 \\
FeI        &            4021.85 \\
TiII       &            4025.14 \\
TiII       &            4028.34 \\ 
MnI        &            4030.80 \\
FeI        &            4045.81 \\
TiII       &            4053.84 \\
FeI        &            4063.60 \\
FeI        &            4067.27 \\
FeI        &            4071.74 \\ 
CrI & 4077.67 \\
H$\delta$  &            4101.70 \\  
FeI        &            4107.50 \\ 
FeI        &            4118.89 \\  
FeII       &            4122.61 \\  
FeII       &            4128.75 \\  
FeI        &            4132.05 \\  
FeI        &            4134.68 \\  
FeI        &            4143.87 \\  
FeI        &            4147.68 \\ 
FeI        &            4154.81 \\ 
FeI        &            4161.62 \\  
FeI        &            4163.66 \\  
MgI        &            4167.27 \\  
FeII       &            4173.46 \\   
FeII       &            4178.85 \\   
FeI        &            4191.43 \\   
FeI        &            4202.02 \\  
FeI        &            4216.18 \\  
CaI-r      &            4226.73 \\  
FeII       &            4233.16 \\  
FeI        &            4250.79 \\  
CrI        &            4254.35 \\  
FeII       &            4258.16 \\  
FeI        &            4260.49 \\    
FeI        &            4271.77 \\   
FeII       &            4273.33 \\
CrI        &            4274.81 \\
FeI        &            4282.40 \\
TiII       &            4290.23 \\
FeI        &            4294.10 \\
FeI/TiII/FeII &            4294.10 \\
FeII       &            4296.57 \\
TiII       &            4300.05 \\
TiII       &            4301.91 \\
FeII       &            4303.18 \\
TiII       &            4307.88 \\
TiII       &            4312.86 \\
ScII       &            4314.08 \\
TiII       &            4316.80 \\
TiII       &            4320.96 \\
FeI        &            4325.77 \\ 
TiII       &            4330.62 \\
H$\gamma$  &            4340.50 \\
FeII       &            4351.77 \\
TiII       &            4367.63 \\
FeI        &            4369.78 \\
FeI        &            4375.93 \\ 
FeI        &            4383.54 \\
FeII       &            4385.39 \\
TiII       &            4395.04 \\
TiII       &            4399.76 \\
FeI        &            4404.75 \\
FeI        &            4408.43 \\
FeI        &            4415.11 \\
FeII       &            4416.82 \\
FeI        &            4427.31 \\
FeI        &            4430.62 \\ 
FeI        &            4442.33 \\
TiII       &            4443.80 \\
FeI        &            4447.72 \\
TiII       &            4450.49 \\
FeI        &            4459.12 \\
FeI        &            4461.65 \\
TiII       &            4468.49 \\
HeI        &            4471.57 \\
FeI        &            4482.17 \\
FeII       &            4489.20 \\
FeII       &            4491.39 \\ 
FeI        &            4494.56 \\
TiII       &            4501.27 \\
FeII       &            4508.28 \\
FeII       &            4515.32 \\
FeII       &            4520.22 \\
FeII       &            4522.63 \\
FeI        &            4528.61 \\
FeI        &            4531.15 \\
FeII       &            4534.00 \\
FeII       &            4541.52 \\
FeII       &            4549.51 \\ 
FeII       &            4555.89 \\
CrII       &            4558.60 \\
TiII       &            4563.77 \\
MgI        &            4571.10 \\
FeII       &            4583.83 \\
CrII       &            4588.20 \\
TiII       &            4589.92 \\
FeI        &            4592.65 \\
FeI        &            4602.94 \\ 
CrII       &            4616.59 \\
FeI/CrII   &            4618.77 \\
FeII       &            4620.52 \\
FeII/TiI   &            4629.34 \\
CrII       &            4634.07 \\
FeI        &            4654.50 \\
TiII       &            4657.21 \\
FeII       &            4666.76 \\ 
FeII       &            4670.18 \\
FeI        &            4691.41 \\
MgI        &            4703.01 \\ 
FeII       &            4731.45 \\
FeI        &            4733.59 \\
CrI        &            4764.63 \\
TiII       &            4779.98 \\
FeI        &            4786.81 \\
FeI        &            4789.65 \\ 
FeII      &            5018.43 \\
TiI       &            5036.47 \\
TiI       &            5038.4  \\
FeI       &            5041.34 \\
FeI       &            5049.82 \\
FeI       &            5051.63 \\
FeI & 4991.28 \\
FeI &  4993.70 \\
FeI & 5006.12 \\
FeI &  5012.08 \\
FeI  & 5030.78 \\ 
FeI        &            5065.19 \\
FeI        &            5068.77 \\
FeI        &            5072.67 \\
FeI        &            5079.74 \\
FeI        &            5083.34 \\
YII        &            5087.44 \\  
FeI        &            5098.70 \\
FeII       &            5100.66 \\
FeI:       &            5107.45 \\
FeI        &            5110.41 \\
FeII       &            5120.35 \\
FeI        &            5123.72 \\
FeI        &            5127.36 \\
TiII       &            5129.15 \\
FeII       &            5132.67 \\ 
NiI        &            5137.08 \\
FeI        &            5139.25 \\
FeI        &            5142.93 \\
FeI        &            5146.30 \\
FeI        &            5151.91 \\
FeI        &            5154.10 \\
MgI        &            5167.32 \\
FeII       &            5169.03 \\
MgI        &            5172.68 \\
Mg I       &            5183.60 \\
NiI        &            5184.56 \\
TiII       &            5188.70 \\
FeI        &            5191.46 \\
FeI        &            5194.94 \\
FeII       &            5197.58 \\
FeI        &            5202.34 \\
CrI        &            5204.52 \\
FeI        &            5208.59 \\ 
FeI       &            5216.27 \\
FeI       &            5226.86 \\
FeII      &            5234.62 \\
CrII      &            5237.33 \\
FeI       &            5247.05 \\
FeI       &            5250.21 \\
CrI  & 5255.12 \\
FeI       &            5269.54 \\
FeI       &            5273.37 \\ 
FeII      &            5276.00    \\   
FeII      &            5284.11 \\ 
CrI        &            5296.69 \\
CrI        &            5298.01 \\
CrI        &            5300.70 \\
FeI        &            5302.31 \\
FeII       &            5306.18 \\
FeI        &            5307.37 \\
FeII       &            5316.61 \\
FeI        &            5324.18 \\
FeI        &            5326.14 \\
FeI        &            5328.04 \\
FeI        &            5332.90 \\
TiII       &            5336.81 \\
FeI        &            5341.02 \\
CrII       &            5346.08 \\
CrI        &            5348.31 \\ 
FeI        &            5353.38 \\
FeII       &            5362.86 \\ 
FeI        &            5371.49 \\
TiII       &            5381.13 \\
FeI        &            5383.37 \\
FeI        &            5393.24 \\
MnI        &            5394.70 \\
FeI        &            5397.20 \\
FeI        &            5405.77 \\
MnI        &            5407.42 \\
CrI        &            5409.77 \\
FeI        &            5415.20 \\
TiII       &            5418.77 \\
CrII       &            5420.92 \\
FeII       &            5425.26 \\
FeI        &            5429.70 \\
FeI        &            5432.95 \\
FeI        &            5434.52 \\ 
FeI        &            5446.87 \\
FeI        &            5455.61 \\
FeI        &            5476.57 \\
FeI        &            5497.52 \\
FeI        &            5501.47 \\
FeI        &            5506.78 \\
FeI        &            5512.26 \\
TiI        &            5514.30 \\ 
MgI        &            5528.40 \\
FeII       &            5534.85 \\
FeI        &            5569.62 \\
FeI        &            5572.84 \\
{[OI]}      &            5577.34 \\
FeI        &            5586.87 \\
NiI        &            5592.28 \\
CaI        &            5598.49 \\
FeI        &            5602.94 \\
FeI        &            5615.64 \\ 
FeI       &            5624.54 \\
FeII     &             5627.48  \\  
FeI       &            5641.43 \\
FeI  &            5658.81    \\ 
FeI       &            5662.94 \\
SiI       &            5684.48 \\
FeI       &            5701.54 \\ 
FeI/NiI       &            5709.38 \\   
FeI       &            5712.13 \\ 
FeI       &            5853.68 \\ 
HeI &  5875.65 \\  
NaD       &            5895.94 \\
FeI & 5956.70  \\ 
FeII      &            5991.38 \\
FeI       &            6065.48 \\
FeII      &            6084.11 \\ 
NiI        &            6108.12 \\
FeII       &            6113.33 \\
CaI        &            6122.22 \\
FeI        &            6136.62 \\
FeI        &            6137.83 \\
FeI        &            6141.73 \\
FeII       &            6147.73 \\
FeI        &            6148.65 \\
FeII       &            6149.24 \\
CaI        &            6162.17 \\
FeI        &            6173.34 \\
FeI/NI     &            6191.56 \\ 
FeI        &            6213.43 \\
FeI        &            6219.29 \\
FeI        &            6230.72 \\
FeII       &            6238.38 \\
FeII       &            6247.56 \\
FeI        &            6252.56 \\
FeI        &            6254.26 \\
FeI/NI     &            6256.36 \\ 
TiI        &            6261.10 \\
FeI        &            6265.13 \\
FeI        &            6270.23 \\
FeI        &            6280.62 \\
{[OI]}       &            6300.30 \\ 
FeI       &            6318.02 \\
FeI       &            6335.33 \\
FeI       &            6358.63 \\
{[OI]}      &            6363.78 \\
FeII      &            6369.46 \\
FeI       &            6393.6  \\
FeI       &            6400    \\
FeII      &            6416.9  \\ 
FeII       &            6546.38 \\
H$\alpha$  &            6562.80 \\
TiI        &            6592.92 \\
FeI        &            6609.12 \\
NiI        &            6643.65 \\ 
HeI       &            6678.15 \\
CaI        &            6717.69 \\
{[SII]}       &            6730.816 \\
FeI       &            6750.15 \\
NiI       &            6767.77 \\ 
FeI        &            6945.20 \\
FeI        &            6978.85 \\ 
FeI        &            7222.90 \\ 
FeII       &            7224.48 \\
SiI        &            7289.25 \\
NiI        &            7291.45 \\
{[CaII]}    &            7291.47 \\
FeII        &            7308.00 \\ 
Unidentified        &     7310.33 \\
Unidentified        &            7320.94 \\
{[CaII]}     &            7323.89 \\ 
FeI        &            7386.34 \\
MgI        &            7387.69 \\
SiI        &            7415.36 \\
SiI        &            7415.95 \\
FeII       &            7449.34 \\
FeII       &            7462.38 \\
FeII       &            7479.69 \\ 
FeII       &            7533.37 \\
NiI        &            7555.60 \\
FeI        &            7583.80 \\ 
KI        &            7698.96 \\
FeII       &            7711.73 \\ 
FeI:  & 7714.6061 \\ 
FeI       &            7748.28 \\ 
OI & 7771.94 \\  
FeI       &            7780.5568 \\ 
NiI       &        7788.94  \\ 
FeI       &           7912.87 \\ 
FeI       &            7937.14\\ 
\end{longtable}